\documentclass[twocolumn,prb,showpacs,superscriptaddress]{revtex4}
\usepackage{graphicx}
\usepackage{amsmath}
\usepackage{amssymb}
\newcommand{\figwidth}{0.95\columnwidth}
\newcommand{\largefigwidth}{0.85\columnwidth}
\newcommand{\Tr}{\operatorname{Tr}}

\newcommand{\pd}{\phantom{\dagger}}
\newcommand{\sign}{\operatorname{sign}}
\renewcommand{\Re}{\operatorname{Re}}

\begin{document}
\title{Quantum quench dynamics of some  exactly solvable models in one dimension}
\author{A.~Iucci}
\affiliation{Instituto de F\'{\i}sica la Plata (IFLP) - CONICET and Departamento de F\'{\i}sica,\\
Universidad Nacional de La Plata, CC 67, 1900 La Plata, Argentina}
\affiliation{DPMC-MaNEP, University of Geneva, 24 Quai Ernest Ansermet CH-1211 Geneva 4, Switzerland}
\affiliation{Donostia International Physics Center (DIPC), Manuel de Lardiz\'abal 4, 20018 San Sebasti\'an, Spain}
\author{M.~A.~Cazalilla}
\affiliation{Centro de F\'{\i}sica de Materiales (CFM). Centro Mixto CSIC-UPV/EHU. Edificio Korta, Avenida de Tolosa
72, 20018 San Sebasti\'an,  Spain}
\affiliation{Donostia International Physics Center (DIPC), Manuel de Lardiz\'abal 4, 20018 San Sebasti\'an, Spain}
\begin{abstract}
 The dynamics of the Luttinger model and the sine-Gordon model (at the Luther-Emery point as
 well as in the semiclassical approximation) after a quantum quench is studied. We compute in detail one and two-point
correlation functions for different types of quenches: from a non-interacting to an interacting Luttinger
model and vice-versa, and from the gapped to the gapless phase of the sine-Gordon model and vice-versa.
A progressive destruction of the Fermi gas features in the momentum distribution is found in the case of
a quench into an interacting state in the Luttinger model. The critical exponents for spatial correlations
are also found to be different from their equilibrium  values.
Correlations following a  quench of the sine-Gordon model from the gapped to the gapless phase
are found in agreement with the  predictions of Calabrese and Cardy [Phys. Rev. Lett.  {\bf 96} 136801  (2006)].
However, correlations following a  quench from the gapped to the gapless phase at the Luther-Emery and
the semi-classical limit exhibit a somewhat different behavior, which may indicate a break-down of the semiclassical
approximation or a qualitative change in the behavior of correlations as one moves away from the Luther-Emergy point.
In all cases, we find that the correlations at infinite times after the quench  are
well described by a generalized Gibbs ensemble [M. Rigol \emph{et al.} Phys. Rev. Lett. {\bf 98},
050405 (2007)], which assigns a momentum dependent temperature to each eigenmode.
\end{abstract}
\maketitle
\section{Introduction}

Most of the theoretical effort in the field of strongly correlated
quantum systems over the past few decades has focused on
understanding the equilibrium properties of these fascinating
systems. For instance, achieving a complete understanding of the
phase diagram of rather ``simple'' models like the
two-dimensional fermionic Hubbard model still remains a huge
challenge. Nevertheless, however important these endeavors are,
understanding the phase diagram and the equilibrium properties of the phases
of strongly correlated systems will not certainly exhaust the possibilities for finding new and
surprising phenomena in these complex systems, especially out of 
equilibrium.

In classical systems, the existence of steady  states out of equilibrium 
is well known. Very often, however, the properties of such  states
have very little to do with the equilibrium properties of the systems
where they occur. Moreover, also very often  their
existence cannot be inferred from
any previous knowledge about the  equilibrium phase diagram: They
are \emph{emergent} phenomena. One good example is provided by the
appearance of  Rayleigh-Benard  convection cells  when fluid
layer is driven out of equilibrium by the application of a
temperature gradient. Although
dissipation plays an important role in the formation of those classical
non-equilibrium  states, one may also wonder if also 
non-quilibrium steady states can appear
when quantum systems are driven out of equilibrium. Differently from classical
systems, dissipation in
quantum systems causes decoherence, which  usually destroys
any quantum interference effects that could lead to quantum
non-equilibrium states without classical analog. However, it is known that
decoherence, due to  coupling with the environment, is always present in
most quantum many-body systems for large enough collections of
particles.  This may explain why the study of non-equilibrium
phenomena in quantum many-body systems has been  regarded, until
very recently, as a subject of mostly academic interest.

The recent availability  of highly controllable systems such as
mesoscopic heterostructures (\emph{e.g.} quantum dots) and
especially ultra-cold atomic gases has finally provided the largely lacking
experimental motivation for the study of non-equilibrium
phenomena, leading to an explosion of theoretical
activity.\cite{altman_quench_2002,sengupta_quench_QCP,barankov_%
dynamical_projection,yuzbashyan_BCS_quench,kollath_density_waves,%
altman_projection_feshbach_molecules,ruschhaupt_quench_momentum_%
interference,yuzbashyan_quench_fermions_order_parameter,cazalilla_%
quench_LL,rigol_generalized_gibbs_hcbosons,perfetto_quench_coupled_LL, %
calabrese_quench_CFT,Rigol_hc_noneq2,manmana_quench_spinless_fermions_nnn,
calabrese_quench_CFT_long,kollath_quench_BH,Gritsev_spectroscopy_07,%
eckstein_generalized_gibbs_FK,eckstein_generalized_gibbs_hubbard,%
kehrein_quench_hubbard,eisert_crap1,eisert_crap2,rigol_noninteg,Manmana_long,%
degrandi_adiabatic_quench,Reimann08,faribault_fermion_pairing_model,Patane08,Patane08b,%
Rossini08,Barmettler08,sen_nonlinear_quench_QCP,sen_2,sen_3,sen_4,sen_5,sen_6,sen_7,roux_time_dependent_lanczos}
Cold atomic gases are especially
interesting because they are very weakly coupled to the
environment, thus remaining fully quantum coherent for fairly long
times (compared to the typical duration of an experiment). At the
same time, it is relatively easy to measure the \emph{coherent}
evolution in time of observables such as the density or the
momentum distribution. Thus,  theorists can  now begin to
pose questions such as: Assuming that a many-body system is prepared in a
given initial state that is not an eigenstate of  the Hamiltonian,
how will it evolve in time? And, more specifically,
will it reach a stationary or quasi-stationary state? If so,
what will be the properties of such a state?
How much memory will the system retain of its initial conditions?

From another point of view, the problem described in the previous
paragraph can be formulated as the study of the response of a
system to a sudden perturbation in which the Hamiltonian is
changed over a time scale much
shorter than any other characteristic time scales of the system.
In what follows, we shall refer to this type  of experiment as a
\emph{quantum quench}. Quantum quenches are also of particular
interest to the `quantum engineering' program for cold atomic
gases.\cite{quantum_engineering}
The reason is that, if we intend to use these
highly tunable and controllable systems as
quantum simulators of models  of many-body
Physics (such as the 2D fermionic Hubbard model mentioned above),
it is  utterly important to understand  to what extent
the final state of the quantum simulator depends on the  state in
which it was initially prepared.  In particular, one is interested
in finding out whether the observables in the final state state
can be obtained from a standard statistical ensemble (say, the
microcanonical, or the canonical ensemble at an effective
temperature). If this is so, one would speak of
\emph{thermalization}. If this does not happen, then how much
memory does  the system retain about its initial state beyond the
average energy  $E =  \langle H \rangle$?

We would like to
emphasize that the above questions are not a merely academic.
Indeed, cold atomic systems allow for the study of non adiabatic dynamics
when the system is driven between two quantum phases such as
a superfluid and a Mott insulator.
\cite{greiner_transition_superfluid_mott,greiner_fast_tunnability}
Also, in a recent experiment, \cite{kinoshita_non_thermalization}
it has been shown that a  faithful realization of the Lieb-Liniger
model \cite{lieb_liniger_model} exhibits absence of
thermalization. In other words, when prepared in an
non-equilibrium state, the experimental system reached a steady
state that cannot be described by any  of the `standard' ensembles of
Statistical Mechanics. This absence of thermalization seems to be
a consequence of the \emph{integrability} of the Lieb-Liniger
model, that is, the existence of an infinite number of
independently conserved quantities.  This conclusion
was backed by the theoretical analysis of Rigol and
coworkers,\cite{rigol_generalized_gibbs_hcbosons} who
observed that the non-equilibrium dynamics of an integrable
system is highly constrained. Thus, based
on numerical simulations for the Tonks-Girardeau limit
of the Lieb-Liniger model, these authors conjectured
that the long-time values of  some observables should converge to those
obtained from a Generalized Gibbs ensemble, which
can be constructed from a maximum entropy
approach.~\cite{JaynesI, JaynesII,rigol_generalized_gibbs_hcbosons}
The conjecture was first analytically confirmed  by analyzing
an interaction quench in the Luttinger
model by one of us.~\cite{cazalilla_quench_LL}
Later, it has been also found true  in other
integrable models: Cardy and Calabresse studied
a quench in a Harmonic chain,~\cite{calabrese_quench_CFT_long}
Eckstein and Kollar analyzed
the Falikov-Kimball model in
infinite dimensions,\cite{eckstein_generalized_gibbs_FK} and
the $1/r$  Hubbard model in one dimension.~\cite{eckstein_generalized_gibbs_hubbard}
Moeckel and Kehrein~\cite{kehrein_quench_hubbard}
studied an interaction quench in the Hubbard model
in infinite dimensions  by a flow equation method,
and found that the system reaches
an intermediate non-thermal state. Finally, recent numerical
studies also have suggested  that lack of thermalization may even
persist in the absence of integrability in one-dimensional
systems,\cite{manmana_quench_spinless_fermions_nnn}
or that  it  may occur only certain parameter regimes of
non-integrable models.\cite{kollath_quench_BH}

However, it can be expected  that,~\cite{Reimann08}
for a rather general choice of the
initial state, along with a situation where there are few
conserved quantities, the system will lose memory of most
of the details of the initial state and, after it reaches a
steady state, the expectation of
most experimentally accessible observables such as
the particle density or the momentum distribution, will
look essentially  as those obtained from a standard
thermal ensemble.\footnote{Implicit in this
discussion it is the fact we are attempting at a description of the system as a
whole, not separating its degrees of freedom into a `system'  and
a `reservoir'. We believe that this point of view may be  the more
appropriate when discussing   cold atomic systems, given that they
are very weakly coupled to the environment.}
Indeed, this is what seems to
be observed  in the vast majority of the experiments with
cold atomic gases. However, for experimental many-body
systems in general, it is hard to quantify whether this will
be always the case. We should take into account
that (except perhaps in the case of cold atomic gases)
the exact form of the quantum Hamiltonian is frequently
not known with accuracy. And even when it is known, it is not always
possible to tell a priori whether the system
is integrable or even if it has other conserved quantities besides
the ones assumed by the standard thermodynamic
ensembles.  As a possible experimental check,
we can say that,  provided the final result is largely
independent of the  particular
details of the preparation of the initial state and
provided that some simple observables can be computed
assuming that the final steady state is thermal,
we can say that thermalization has occurred.
Indeed,  some recent numerical evidence,\cite{rigol_noninteg}
supplemented by the extension to many-particle systems
of a conjecture known as `eigenstate thermalization hypothesis'
(first introduced in the context of quantum chaos\cite{Srednicki_eigenstate}),
seems to indicate that  lack of integrability will in general lead
to thermalization (in the sense defined above). Indeed, Reimann~\cite{Reimann08} has
recently analytically demonstrated that, under realistic experimental conditions,
equilibration will be observed in an isolated system that has been initially
prepared in an non-equilibrium mixed state.
Nevertheless, even of the issue of thermalization for non-integrable systems
may have been finally settled, other questions such as the details of the
transition from the integrable case (which thermalized to a
generalized Gibbs ensemble) to the non integrable
case (which thermalizes to the standard microcanonical,
or for large enough systems, Gibbs ensembles) are questions
that are still far from being completely understood.~\footnote{This question
is also related to the problems concerning
the applicability of the maximum entropy approach~\cite{JaynesI,JaynesII}
to Statistical Mechanics. See for instance the critique
by S. K. Ma in Ref.~\onlinecite{ma_stat_mech},. However, the
maximum entropy approach is advocated by R. Balliant.~\cite{balian_stat_mech}}

In this article, we will not try to answer the difficult questions
posed in the previous paragraph. Instead, we focus on analyzing
the quench dynamics of two relatively well-known
one dimensional models:  the Luttinger model (for which a brief
account of the results has  been already published elsewhere\cite{cazalilla_quench_LL})
and the sine-Gordon model. For the latter, we
present results in two solvable limits in which the Hamiltonian
can be reduced to a quadratic form of creation and destruction
operators.  The  long time behavior of the single particle
Green's function after a quench for   general
quadratic Hamiltonians of this form has been recently studied
by Barthel and Scholw\"ock.\cite{schollwoeck_quench}
These authors provided some general
conditions for the appearance of dephasing and
steady non-thermal states. This question as been
also taken up recently by Kollar and
Eckstein.~\cite{eckstein_generalized_gibbs_hubbard}
However, since the models taken up in this article
may be relevant  for the experiments with cold atomic gases
(see Sect.~\ref{sec:exp}) or numerical simulations,
it is important to obtain analytical results for them.
The simplicity of these models also allows us
to test in detail a number of  general
results.~\cite{calabrese_quench_CFT,schollwoeck_quench}

The rest of this article is organized as follows. In
Sect.~\ref{sec:quadratic} we describe the how the
dynamics after a quantum quench can be obtained as a time-dependent canonical transformation for
fairly general quadratic Hamiltonians. The solution
is used extensively in Sects.~\ref{sec:LM} and~\ref{sec:sG}
to obtain the evolution of observables and correlation
functions of the Luttinger and sine-Gordon models.
For the Luttinger model we consider the case
where the interaction between the fermions is suddenly switched
on (for which some of the results where briefly reported
in Ref.~\onlinecite{cazalilla_quench_LL}) and the reverse
situation, when the interaction in suddenly switched off.
For the sine-Gordon model, we study the situations when
the system is quenched between the gapped and the gapless
phases of the model, and viceversa. We show that in
the latter case  (\emph{i.e.} when the system is quenched
from the gapless to the gapped phase), the results
obtained at the Luther-Emery point (where the model
can be \emph{exactly} mapped to a quadratic fermionic
Hamiltonian) and in the so-called semiclassical approximation
are qualitatively different. The origin of this difference is
still not well understood, and we cannot discard that it
is indeed an artifact of the quasi-classical approximation
(which, however, yields results in agreement with those
obtained at the Luther-Emery point for the case when the
system is quenched form the gapped to the gapless phase).
In Sect.~\ref{sec:generalized} we show the long time behavior
of some of the correlation functions in these models can be
obtained from a generalized Gibbs ensemble
(along with instances where it fails).
The experimental relevance of our results is
briefly discussed in Sect.~\ref{sec:exp}, along with
other conclusions of this work. Finally, the details of
some of the most lengthy calculations are provided
in the appendices.

\section{Quadratic Hamiltonians}\label{sec:quadratic}

As an illustrative calculation, let us first study the case of a
quantum quench in a model described by a quadratic Hamiltonian:
\begin{multline}
H(t) =\sum_{q} \hbar \left[  \omega_0(q) + m(q,t)
\right] b^\dag(q) b(q) \\
\quad + \frac{1}{2} \sum_{q} \hbar g(q,t) \left[ b(q) b( -q) +
b^\dag(q) b^\dag(-q)\right], \label{eq:genham}
\end{multline}
where $[ b(q), b^\dag(q')] = \delta_{q,q'}$, commuting otherwise.
We will assume that the quench takes place at $t = 0$, so that,
within the sudden approximation, the system is described by
$H_\text{i} = H(t \leq 0)$ for $t < 0$ and by $H_\text{f} = H(t >
0)$ for $t  >  0$.  Furthermore, in order to simplify the analysis, we
assume that $m(q,t \leqslant 0) = g(q,t \leqslant 0) = 0$, and
$m(q,t>0) = m(q)$ and $g(q,t) = g(q)$. Notice that the initial
Hamiltonian is diagonal in the $b$-operators:
\begin{equation}
H_\text{i}=H_0\equiv\sum_q
\hbar\omega_0(q)b^\dagger(q)b(q).\label{eq:hamiltonian_H_0}
\end{equation}
In order  to obtain the time evolution of operators  $O =
\mathcal{O}[\{b^\dag(q), b(q) \}]$ after the quench, we recall
that, in the Heisenberg picture, $\mathcal{O}(t > 0) = e^{i
H_\text{f} t/\hbar} \mathcal{O} e^{-i H_\text{f} t/\hbar} =
\mathcal{O}(\{ b(q,t), b^\dag(q,t) \})$, and therefore all that is
needed to solve the above quench problem is to obtain the time
evolution of $b(q)$ for $t > 0$. For Hamiltonians
like~(\ref{eq:genham}) this can be done exactly because
$H_\text{f}  = H(t > 0)$ can be diagonalized by means of the
canonical Bogoliubov (``squeezing'') transformation:
\begin{align}
a(q) & = \cosh \beta(q) \, b(q) + \sinh \beta(q) \, b^\dag(-q).
\label{eq:bogol}
\end{align}
Upon choosing
\begin{equation}
\tanh 2\beta(q) = \frac{g(q)}{\omega_0(q) +
m(q)},\label{eq:bogol_beta}
\end{equation}
the Hamiltonian at $t > 0$ is rendered diagonal:
\begin{equation}
H_\text{f} = H\equiv E_0 +  \sum_{q} \hbar \omega(q) \,
a^{\dag}(q) a(q),\label{eq:ham_diag}
\end{equation}
where $E_0$ is the energy of the ground state of $H$ (relative to
the ground state energy of $H_0$) and
\begin{equation}
\omega(q) = \sqrt{\left[ \omega_0(q) + m(q)\right]^2 - [g(q)]^2}
\end{equation}
the dispersion of the excitations about the ground state of
$H_\text{f}$. The evolution of the $a(q)$ is given by $a(q,t) =
e^{iH_\text{f} t/\hbar} a(q) e^{-iH_\text{f} t/\hbar} = e^{-i
\omega(q) t} a(q)$. By application of a direct and reverse
Bogoliubov transformation, one can obtain the time evolution of
$b(q)$:
\begin{equation}
b(q,t) = f(q,t)\, b(q) + g^*(q,t) \,
b^\dag(-q),\label{eq:solution_time}
\end{equation}
where
\begin{align}
f(q,t) &= \cos \omega(q) t  - i \sin \omega(q) t \, \cosh
2\beta(q),
\label{eq:fqt}\\
g(q,t) &= i \sin \omega(q) t \, \sinh 2\beta(q).\label{eq:gqt}
\end{align}
It is easy to check that (\ref{eq:solution_time}) obeys the
initial condition, $b(q,t =0) = b(q)$, and also respects the
equal-time commutation rules,
\begin{align}
[b(q,t), b(q',t)] &= \big( f(q,t) g^{*}(q,t)
\notag \\
&\quad - g^*(q,t) f(q,t)\big)  \delta_{q,-q} = 0,\\
[b(q,t), b^\dag(q',t) ] &=  \left(  |f(q,t)|^2 - |g(q,t)|^2\right)
\delta_{q,q'} \nonumber \\
&= \delta_{q,q'}
\end{align}
Thus, a quantum quench described by a quadratic Hamiltonian can be
solved by means of a time-dependent canonical transformation.

When the quench is reversed, \emph{i.e.} when the case with $m(q,t
\geqslant 0) = g(q,t \geqslant 0) = 0$, and $m(q,t<0) = m(q)$ and
$g(q,t<0) = g(q)$ is considered, the roles played by the initial
and final Hamiltonians are also reversed: the \emph{final} Hamiltonian
is now diagonal in the $b$'s, $H_\text{f}=H_0$, whereas the
transformation of Eq. (\ref{eq:bogol}) renders the \emph{initial}
Hamiltonian, $H_\text{i}=H$, diagonal. Therefore, in this case the
evolution of the $b$-operators is trivial:
$H_\text{f}$: $b(q,t)=e^{-i\omega_0(q)t}$, whereas the evolution of
the $a$'s is given by
\begin{equation}
a(q,t) = f_0(q,t)\, a(q) + g_0^\ast(q,t) \, a^\dag(-q),
\end{equation}
where
\begin{align}
f_0(q,t) &= \cos \omega_0(q) t  - i \sin \omega_0(q) t \, \cosh
2\beta(q),
\label{eq:f0qt}\\
g_0(q,t) &= -i \sin \omega_0(q) t \, \sinh 2\beta(q).\label{eq:g0qt}
\end{align}

\section{The Luttinger model}\label{sec:LM}

The Luttinger model (LM) is a one-dimensional (1D) system of interacting fermions
with linear dispersion. It was first described by Luttinger \cite{luttinger_model}
but its complete solution was later obtained
by Mattis and Lieb,\cite{mattis_lieb_luttinger_model}
who showed that the elementary excitations of the system are not
fermionic quasi-particles. Instead, they introduced a set of bosonic operators
describing collective density modes (phonons) of the system, which are the true
elementary low-energy excitations  of the model. The methods of Mattis and
Lieb bear strong resemblance to the early work of Tomonaga~\cite{Tomonaga_1D_electron_gas}
on the one-dimensional electron gas. Extending the work of Tomonaga, Mattis and Lieb,
Luther and Peschel \cite{luther_peschel_correlation_functions}
computed the equilibrium one and two-particle correlation functions,
showing that all correlation functions exhibit (at zero temperature)
a non-universal power-law behavior at long distances, which signals the absence
of long-range order.  Later, Haldane \cite{haldane_exponents_spin_chain,%
haldane_luttinger_liquid,haldane_effective_harmonic_fluid_approach}
conjectured that these properties (\emph{i.e.} collective elementary excitations exhausting
the low-energy part of the spectrum as well as power-law correlations) are
a distinctive features of a large class of gapless interacting one-dimensional systems,
which he termed (Tomonaga-)`Luttinger liquids'. Thus, the Luttinger model can be understood
as a  fixed point of the renormalization-group for a large class of gapless many-body
systems in one dimension. Therefore, the thermal equilibrium properties of many 1D system
are \emph{universal} in the sense that they can be accurately described by the Luttinger model.
However, in this work we shall be concerned
with the non-equilibrium properties of the LM,  and  because non-equilibrum phenomena
can involve highly excited states,  we shall make no claim for universality.
The precise conditions conditions under which the results obtained here apply to
real systems that are the  Tomonaga-Luttinger liquids should be investigated carefully
for each particular system (see also discussion in Sect.~\ref{sec:exp}).

The Hamiltonian of the Luttinger model (LM) can be written as follows:
\begin{align}
H_\text{LM} &=  H_0 + H_2 + H_4,  \label{eq:hlm1}\\
H_0 &= \sum_{p,\alpha = r,l} \hbar v_F p :\psi^{\dag}_{\alpha}(p)
\psi_{\alpha}(p):\,, \label{eq:hlm2}\\
H_2 &=  \frac{2\pi \hbar}{L}\sum_{q} g_2(q) :J_r(q) J_l(q):\,, \label{eq:hlm3} \\
H_4 &= \frac{\pi\hbar}{L} \sum_{q, \alpha = r,l} g_4(q)
\, :J_{\alpha}(q) J_{\alpha}(-q): \,.\label{eq:hlm4}
\end{align}
The index $\alpha = r,l$ refers to the \emph{chirality} of the
fermion species, which can be either right ($r$) or left ($l$)
moving; the symbol $:\ldots:$ stands for normal ordering prescription
for fermionic operators, which is needed to remove from the
expectation values the infinite contributions arising from the
ground state.\cite{haldane_luttinger_liquid}
The latter is a Dirac sea, namely state where all
single-particle fermion levels with $p < 0$ are occupied for both
chiralities. This defines a stable ground state (at least at the
non-interacting level), which will be denoted by $| 0\rangle$.

\subsection{Bosonization solution of the LM}

 In this section we briefly review the solution of the
LM. The Hamiltonian in Eqs.~(\ref{eq:hlm1}) to (\ref{eq:hlm4})
can be written as a quadratic Hamiltonian in
terms of a set of bosonic
operators.\cite{mattis_lieb_luttinger_model} First note that the
density (current) operators $J_{\alpha}(q) = \sum_{p}
:\psi^{\dag}_{\alpha}(p+q) \psi_{\alpha}(p):$ obey the following
commutation rules:
\begin{align}
\left[J_{\alpha}(q),J_{\beta}(q') \right] = \left(\frac{qL}{2\pi}
\right) \delta_{q+q',0} \delta_{\alpha\beta},
\end{align}
which can be transformed into the Heisenberg algebra of bosonic
operators by introducing:
\begin{equation}
b(q) = - i \left(\frac{2\pi}{|q|L}\right)^{1/2} \left[ \vartheta(q)
J_r(-q) - \vartheta(-q) J_l(q) \right].\label{eq:bosons}
\end{equation}
where $\vartheta(q)$ is the step function. Note that the $q = 0$
components (sometimes calle zero modes) require a separate treatment since
$J_{\alpha}(0) = N_{\alpha}$ is the deviation (relative to the ground state)
in the number of fermions of chirality $\alpha = r, l$.  Rather than working
with $N_r$ and $N_l$, it is convenient to introduce:
\begin{equation}
N = N_r + N_l \quad J = N_r - N_l,
\end{equation}
which, since $N_r$ and $N_l$ are integers, must obey the following
selection rule $(-1)^{N} = (-1)^J$ when the Fermi fields obey anti-periodic boundary
conditions ($\psi_{\alpha}(x + L) = -\psi_{\alpha}(x)$ ($L$ is the length of the system),
and therefore  $\psi_{\alpha}(x) = L^{-1/2}\sum_{p} e^{- a_0 |p|} e^{i p
x} \psi_{\alpha}(p)$, with $p = 2(n-\frac{1}{2})\pi/L$, $n$ being an
integer, and  $a_0 \to
0^{+}$).\cite{haldane_luttinger_liquid}

The Hamiltonian $H_\text{LM}$ can be expressed in terms of the bosonic
operators introduced in Eq.~(\ref{eq:bosons}):
\begin{align}
H_0 &= \sum_{q \neq 0} \hbar v_F |q| \,  b^\dag(q) b(q)
+\frac{\hbar \pi v_F}{2L} \left( N^2 + J^2 \right), \\
H_2 &= \frac{1}{2} \sum_{q\neq 0} g_2(q)|q| \, \left[ b(q)
b(-q)
+ b^\dag(q) b^\dag(-q) \right]\notag\\
&\qquad\qquad+\frac{\hbar \pi g_2(0)}{2L} \left( N^2 - J^2 \right),\\
H_4 &= \sum_{q\neq 0} \hbar g_4(q) |q|\,  b^\dag(q) b(q) +
\frac{\hbar \pi g_4(0)}{2L} (N^2 + J^2).
\end{align}

Ignoring, for the moment, the zero mode part (\emph{i.e.} $q = 0$ terms,
involving $J$ and $N$), the above Hamiltonian has
the form of Eq.~(\ref{eq:genham}), with the following identifications: $
\omega_0(q)= v_F |q|$, $m(q,t)  =  g_4(q)|q|$, and $g(q,t)=
g_2(q)|q|$, and it can be therefore be brought into diagonal
form by means of the canonical transformation of Eq.~(\ref{eq:bogol}).
Hence, the Hamiltonian  takes the form of Eq.~(\ref{eq:ham_diag}) with
$\omega(q)=v(q)|q|$, being
$v(q)=\{[v_F+g_4(q)]^2-[g_2(q)]^2\}^{1/2}$, and $q \neq 0$. As to the
zero mode contribution:
\begin{equation}
H_\text{ZM}=\frac{\hbar\pi v_N}{2L}N^2+\frac{\hbar\pi v_J}{2L}J^2,
\end{equation}
where $v_N=v_F+g_4(0)+g_2(0)$ and $v_J=v_F+g_4(0)-g_2(0)$.
This defines the equilibrium solution of the LM. In the following
sections we shall be concerned with the quench dynamics of this model.

\subsection{Suddenly turning-on the interactions}\label{sec:turningon}

Although it is possible to solve the general quench problem
between two interacting versions of the Luttinger model, we shall
focus here  on the cases where the interactions described by
$H_2$ and $H_4$ are suddenly switched on (this section),  and
(next section) switched  off.  Thus, in this section, we shall assume that
 we have made the replacements
$g_{2,4}(q) \to g_{2,4}(q,t) = g_{2,4}(q) \theta(t)$ in Eqs.(\ref{eq:hlm3},\ref{eq:hlm4}).
The Hamiltonian at times $t > 0$
is therefore the interacting LM. In other words, in the notation introduced in
Sect.~\ref{sec:quadratic},
$H_{\rm f}  = H_{0} + H_2 + H_4 = H_{\rm LM}$, whereas the initial
Hamiltonian (for $t \leq 0$) is $H_{\rm i} = H_{0}$. However, we note that,
since both zero modes, $J$ and $N$, are conserved 
 by $H_{\rm i} = H_{0}$ and $H_{\rm f} = H_{\rm LM}$,
their dynamics can be factored out, and
we shall  assume henceforth that we work within the sector of the Hilbert
space where $J = N = 0$ (this sector contains the non-ineracting ground state, $|0\rangle$).
Thus, from now on, we shall omit $H_\text{ZM}$ in all discussions.

As to the initial state, we shall consider that, within the
spirit of the sudden approximation, at $t=0$ the system is
prepared in a Boltzmann ensemble of eigenstates of $H_\text{i} $, at a temperature $T$:
\begin{equation}
\rho_0\equiv\rho(t=0)=Z_0^{-1}e^{-H_\text{i}/T}, \label{eq:init}
\end{equation}
where $Z_0=\Tr e^{-H_\text{i}/T}$ . We shall further assume that the
contact with the reservoir is removed at $t =0$, and that, after
the quench, the system evolves  \emph{unitarily} in isolation.

Whereas Eq. (\ref{eq:solution_time}) defines the solution to the
interaction quench in terms of the modes that annihilate the
initial ground state $|0\rangle$, the solution itself is not
particularly illuminating. To gain some insight into the
properties of the system following the quench, let us compute a
few observables. Amongst them, we  first turn our attention to
the instantaneous momentum distribution, which is the the
Fourier transform of the one-particle density matrix:
\begin{equation}\label{eq_Green_function}
C_{\psi_r}(x,t) = \langle e^{i H_\text{f} t/\hbar}
\psi^{\dag}_r(x) \psi^{\pd}_r(0) e^{-i H_\text{f} t/\hbar}
\rangle_0,
\end{equation}
where $\left\langle\cdots\right\rangle_0$ means that the
expectation value is taken over the ensemble described by $\rho_0$
(cf. Eq.~(\ref{eq:init})). The time dependence of the operators is
dictated by  $H_\text{f}$, as described in Section
\ref{sec:quadratic}. Notice that, since in general $[H_\text{f}, \rho_0]\neq
0$, time translation invariance is broken, and the
above correlation function is explicitly time-dependent.

The time evolution of  $\psi_\alpha(x)$ can be obtained
using the bosonization formula for the field
operator:\cite{luther_peschel_correlation_functions,%
haldane_luttinger_liquid,giamarchi_book_1d}
\begin{equation}\label{eq:bosonization_formula}
\psi_\alpha(x)=\frac{\eta_\alpha e^{is_{\alpha} \pi/4}}{\sqrt{2\pi a}}\, e^{i
s_\alpha\phi_\alpha(x)},
\end{equation}
being  $\eta_r,\eta_l$  two  \emph{Majorana} operators (also known as
Klein factors, which in the present case reduce to two Pauli
matrices) that obey $\{\eta_{\alpha},\eta_{\beta}\} = 2 \delta_{\alpha\beta}$, thus
ensuring the anticommutation of the left- and right-moving Fermi fields;
we have also introduced  the index $s_\alpha=1$ for $\alpha=r$ and
$s_\alpha=-1$ for $\alpha=l$. The bosonic fields
\begin{equation}\label{eq:boson_decomposition}
\phi_\alpha(x)=s_\alpha\varphi_{0\alpha}+\frac{2\pi x}{L}
N_\alpha+\Phi^\dag_\alpha(x)+\Phi^{\pd}_\alpha(x),
\end{equation}
where $[N_\alpha,\varphi_{0\beta}]=i\delta_{\alpha,\beta}$, and,
in terms of Fourier modes,
\begin{equation}\label{eq:boson_modes}
\Phi_\alpha(x) = \lim_{a_0\to 0^{+}} \sum_{q > 0} \left(
\frac{2\pi}{qL} \right)^{1/2} e^{-q a_0 /2} \, e^{i s_\alpha qx}
b(s_\alpha q).
\end{equation}
The details of the calculations of $C_{\psi_{r}}(x,t)$ have been
relegated to the Appendix~\ref{app:single_particle_correlations}.
In this section we will mainly describe the results. However, a number of
remarks about how the  calculations were performed
are in order before proceeding  any further. We first note that
interactions in the Luttinger model are assumed to be
long ranged.~\cite{luttinger_model,mattis_lieb_luttinger_model,haldane_luttinger_liquid}
This can be made explicit in the interaction couplings by writing $g_{2,4}(q) = g_{2,4}(qR_0)$,
where the length scale $R_0 \ll L$ is the interaction range. Thus, just
like system size $L$ plays the role of a cut-off for `infrared' (that is, long wave-length)
divergences, the interaction range, $R_0$
plays the role of an `ultra-violet'  cut-off that regulates
the short-distance divergences of the model. The results given below
were derived assuming a particular  form of the interaction (or
regularization scheme) where the  Bogoliubov parameter (cf. Eq.~\ref{eq:bogol_beta})
is chosen  such that $\sinh 2 \beta(q) = \gamma\, e^{-|q R_0|/2} $.
Furthermore, we replaced $v(q)$ by $v = v(0)$.
Indeed, these approximations are fairly
similar to  the ones used to compute the
time-dependent correlation
functions in equilibrium,\cite{luther_peschel_correlation_functions} given that
the expressions that we obtain for the out-equilibrium correlators are fairly
similar as well  (see Appendix~\ref{app:single_particle_correlations} for details).
This regularization scheme greatly simplifies the calculations while not altering in a significant way
(except  perhaps for pathological cases, like the Coulomb
interaction, where both $v(q)/|q|$ and $\sinh 2 \beta(q)$ are
singular at $q= 0$) the asymptotic  behavior  of the correlators
for distances  much larger than $R_0$.

Returning to the one-particle density matrix defined above in
Eq.~(\ref{eq_Green_function}), we note that it can be
written as  can be written as the product of two factors:
$C_{\psi_r}(x,t)=C^{(0)}_{\psi_r}(x)h_r(x,t)$, where $C^{(0)}_{\psi_r}(x)$ is the
noninteracting one-particle density matrix, and thus $h_r(x,t)$ accounts for
deviations  due to the interactions. Hence, this factorization allows us
to write the instantaneous momentum distribution function as a
convolution:
\begin{equation}\label{eq:Fourier_transform_convolution}
f(p,t)=\int_{-\infty}^{\infty}\frac{dk}{2\pi}\,f^{(0)}(p-k)h_r(k,t),
\end{equation}
where $f^{(0)}(p)= (e^{\hbar v_F p/T} + 1)^{-1}$ is the
Fermi-Dirac distribution, and $h_r(k,t)$ is the Fourier transform
of $h_r(x,t)$ in $x$ variable.

Before presenting the results for the expression of  the one-particle
density matrix as well as the
momentum distribution at finite temperatures,  it is worth
considering the (much simpler looking)  zero-temperature
expression. We first discuss finite-size effects.
For a system of size $L$ we obtain:\cite{cazalilla_quench_LL}
\begin{multline}
C_{\psi_r}(x|L)=C^{(0)}_{\psi_r}(x|L)\left\vert\frac{R_0}{d(x|L)}\right\vert^{\gamma^2}\\
\times\left\vert\frac{d(x-2vt|L)d(x+2vt|L)}{d(2vt|L)d(-2vt|L)}\right\vert^{\gamma^2/2},\label{eq:green_function_LM}
\end{multline}
where $d(z|L)=L|\sin(\pi z/L)|$ is the \emph{cord} function and
$C^{(0)}_{\psi_r}(x|L)=i \left\{2L\sin
\left[\pi(x+ia_0)/L\right]\right\}^{-1}$ ($a_0\to 0^{+}$) the
noninteracting one-particle density matrix. Notice that this
result is valid only asymptotically, that is, for $d(x|L), d(x\pm
2 vt) \gg R_0$. Thus, we see that $C_{\psi_r}(x,t|L)$ is a
periodic function of time with period equal to $\tau_0 = L/2v$. This is in
agreement with the general expectation that correlations in finite-size
systems exhibit time recurrences because its energy spectrum is discrete. Although
the recurrence time generally depends on the details of the energy spectrum,
in the present version of the LM, the spectrum is linear $\omega(q) \simeq v |q|$ and thus,
the energy spacing between (non-degenerate) many-body states is $\Delta_0 \simeq 2\pi \hbar v/L$.
Hence, the recurrence time $\tau_0 \sim 2\pi / \Delta_0$ follows (the extra factor of $\frac{1}{2}$ is explained
by the so-called light-cone effect, see further below).  The periodic behavior exhibited by
the one-particle density matrix~(\ref{eq_Green_function}) implies that, after the quench,
the system will not reach a stationary state with time \emph{independent}
properties  as $t \to +\infty$.  A similar conclusion is reached by analyzing,
\emph{e.g.} the  finite-size version of the
density correlation function,\cite{cazalilla_quench_LL}
\begin{align}
&C_{J_r}(x,t|L) = \langle e^{i H_\text{f} t/\hbar} J_r(x) J_r(0)
e^{-i H_\text{f} t/\hbar} \rangle_0
\nonumber\\
&= - \frac{(1+\gamma^2)/4\pi^2}{\left[d(x|L)\right]^2}
+ \frac{\gamma^2/8\pi^2}{\left[ d(x-2vt|L)\right]^2}
 + \frac{\gamma^2/8\pi^2}{\left[ d(x+2vt|L)\right]^2},\nonumber\\
 \label{eq:dc}
\end{align}
where
\begin{equation}
J_{r}(x) = \frac{1}{L}\sum_{q} e^{i q x} \: J_{r}(q) = :\psi^{\dag}_{r}(x)\psi_r(x):\quad,
\label{eq:currents}
\end{equation}
is the density (also referred to as current) operator in real space.

However,  in the thermodynamic limit $L\to\infty$, the recurrence time
$\tau_0 = L/2v \to +\infty$, and the system does reach a time-independent
steady state.  In this limit,  $d(x|L)\to |x|$, and the asymptotic
form of the single particle density matrix
becomes:\cite{cazalilla_quench_LL}
\begin{equation}
C_{\psi_r}(x,t>0)=G_r^{(0)}(x)\left|\frac{R_0}{x}\right|^{\gamma^2}
\left|\frac{x^2 - (2vt)^2}{(2vt)^2}\right|^{\gamma^2/2}. \label{eq:cpsi}
\end{equation}
Hence,
\begin{equation}
h_r(x,t) = \left|\frac{R_0}{x}\right|^{\gamma^2}
\left|\frac{x^2 - (2vt)^2}{(2vt)^2}\right|^{\gamma^2/2}.
\end{equation}
In order to understand the evolution of the momentum distribution, without actually
having to compute it,  it is very useful to consider the various limits of the above
expression, Eq.~(\ref{eq:cpsi}). First of all, for short times  such that
$2vt\ll |x|$, the function $h_r(x,t)$ is
asymptotically just a time-dependent factor,\cite{cazalilla_quench_LL}
\begin{equation}
h_r(x,t)\simeq
Z(t)=\left(\frac{R_0}{2vt}\right)^{\gamma^2},
\end{equation}
which can be interpreted as a time-dependent  `Landau quasi-particle'
renormalization constant in an effective time-dependent  Fermi-liquid
description of the system. In other words, as time evolves after
the quench, we could imagine that the quasi-particle weight at the Fermi level is
reduced from its initial value ($Z(t=0) = 1$) to $Z(t > 0) < 1$.
At zero temperature, this time-dependent renormalization of the
quasi-particle weight reflects itself in a reduction of the
discontinuity of the momentum
distribution $f(p,t)$ at the Fermi level (which corresponds to
$p = 0$ in our notation). Therefore, at any finite time, the system behaves as
as if it was a Fermi liquid and therefore it keeps memory of the initial state (a
non-interacting Fermi gas).

Yet, for $t\to +\infty$,
$h_r(x,t)$ becomes a power-law:~\cite{cazalilla_quench_LL}
\begin{equation}\label{eq:long_times_correlator_zero_T}
\lim_{t\to\infty}
h_r(x ,t)=\left|\frac{R_0}{x}\right|^{\gamma^2}.
\end{equation}
Interestingly, this time-dependent reduction of the quasi-particle weight after quenching
into the interacting state has been also found in
Ref.~\onlinecite{kehrein_quench_hubbard} when studying a similar quench
in the Hubbard model in the limit of infinite dimensions. In this case, however,
the discontinuity remains finite even for $t \to +\infty$, which is different
from the behavior of the LM, which is known to be a non-Fermi
liquid system  in equilibrium.  These non-fermi-liquid features also persist in the quench
dynamics, as we have found above.

\begin{center}
\begin{figure}
\includegraphics[width=\largefigwidth]{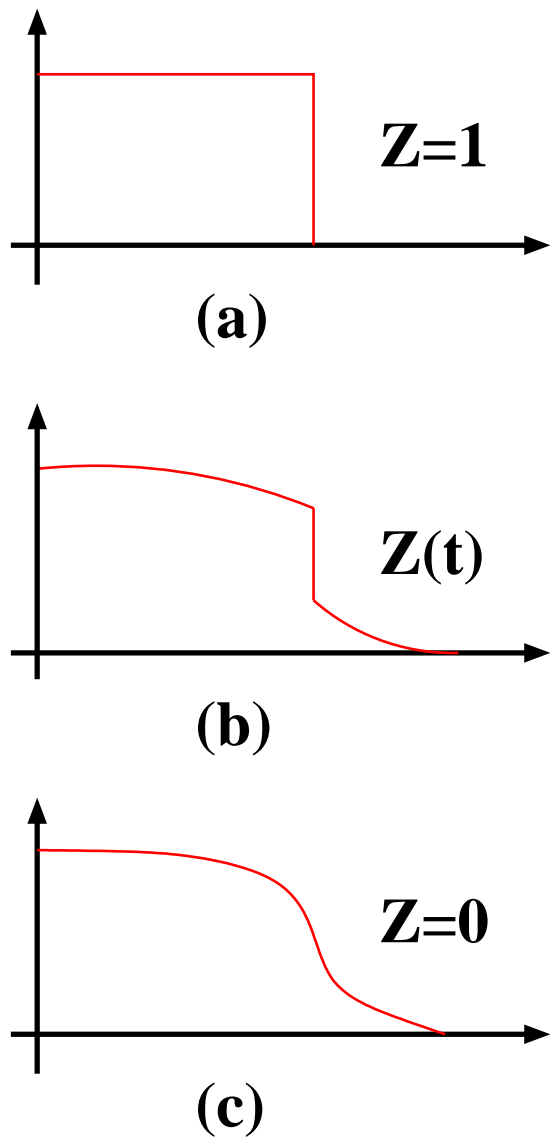}
\caption{Schematic of the time evolution of the momentum distribution $f(p,t)$ at
zero temperature.  (a) At $t=0$, the momentum distribution is that
of non-interacting fermions, with a  discontinuity at the Fermi
level ($p = 0$) $Z=1$. (b) At $t >0$ the discontinuity is reduced
in a power-law fashion $Z(t) \sim t^{-\gamma^2}$. (c) For $t \to
+\infty$ the discontinuity disappears and the momentum
distribution exhibits a power-law singularity close to the Fermi
level $p = 0$, $f(p,t\to +\infty) = \frac{1}{2} - \text{const.}\times
|p|^{\gamma^2}$. However, the exponent characterizing the singularity is
not the equilibrium exponent.}\label{fig:fpt}
\end{figure}
\end{center}

The result of Eq.~(\ref{eq:long_times_correlator_zero_T})
is similar to the zero temperature result in equilibrium. However, the
exponent of $C_{\psi_r}(x,t\to \infty)$ is equal to $1+ \gamma^2$,
and, even for an infinitesimal interaction  (\emph{i.e.} $\gamma \ll
1$), it is always larger than the one that governs the asymptotic ground state (\emph{i.e.} equilibrium)
correlations:\cite{luther_peschel_correlation_functions,haldane_luttinger_liquid}
$\gamma^2  = \sinh^2 2\beta(0) >
\gamma_\text{eq}^2=2\sinh^2\beta(0)$. The reason for the larger exponent
can be understood from two facts: \emph{i}) By the variational theorem, the initial state
(\emph{i.e.} the ground state of the non-interacting Hamiltonian $H_0$), is a
complicated excited state of the system. \emph{ii}) Both $H_0$  and the Hamiltonian that
performs the time-evolution ($H_{\rm f}= H_{\rm LM}$) are critical (\emph{i.e.} scale free, apart from the cut-off $R_0$),
thus, the system is likely to remain critical. Before going further in the discussion of other
results, we  should note that, in the literature on Tomonaga-Luttinger liquids,
it  is customary to introduce the dimensionless parameter $K = e^{-2\beta(0)}$,
in terms of which $\gamma^2 = (K-K^{-1})^2/4$ (whereas $\gamma_\text{eq} =  (K + K^{-1} - 2)/2$).
We shall  use the parameter $K$ in other expressions below.

It is worth emphasizing that the particular evolution of the asymptotic
correlations from Fermi liquid-like at short times to  non-Fermi liquid-like at infinite time
exhibited by the one-particle density matrix, is also found in other correlation functions.
However, the idea that the system 'looks like' an interacting Fermi liquid at any finite $t$
should not be taken too far. In this regard, we should note that
for  $|x| \gg 2 vt$,  the prefactor of the term $\propto (2\pi x)^{-2}$
of the density correlation function (cf. Eq~(\ref{eq:dc})),
which in equilibrium is proportional to the system compressibility,~\cite{giamarchi_book_1d}
remains equal to (minus) unity, which is the value that corresponds
to a non-interacting Fermi gas (in an
interacting Fermi liquid it would deviate from one).
For $t \to +\infty$ the same prefactor becomes $(1 + \gamma^2) > 1$, \emph{i.e.}, which does not
directly reveal the non-Fermi liquid-like behavior. However, other correlation
functions (like the one-particle density matrix discussed above) do.
For instance, consider the following:
\begin{align}
C^m_{\phi}(x,t) &=  \langle e^{2i m \phi(x,t)} e^{-2im\phi(0,t)} \rangle, \label{eq:corr_phi} \\
C^n_{\theta}(x,t) &= \langle e^{in\theta(x,t)} e^{-i n \theta(0,t)} \rangle, \label{eq:corr_theta}
\end{align}
where $\phi(x) = \frac{1}{2}\left[\phi_r(x) + \phi_l(x) \right]$ and $\theta(x) =
\frac{1}{2}\left[\phi_r(x) - \phi_l(x)\right]$.
The spatial derivatives  of $\phi$ and $\theta$ are related to the (total)
density and current density fluctuations, respectively,
in the system.\footnote{In a Luttinger liquid, the $C^m_{\phi}$ correlator
describes the fluctuations of wave number close to
$2m k_F$ (where $k_F$ is the Fermi momentum) of  the density-correlation
function~\cite{haldane_luttinger_liquid,giamarchi_book_1d}. In the presence of a periodic potential
of periodicity equal to $2 m k_F$ the system may become an insulator.~\cite{giamarchi_book_1d}
The power-law behavior exhibited  at zero temperature, and in the thermodynamic limit)
is a consequence  of the gapless spectrum and  the absence of long
range order in the density. In the insulating (\emph{i.e.} gapped) phase, this correlation
function decays to a non-zero constant at long distances, which is a consequence of the existence
of long ranger order in the density at wave  number $2 m k_F$.  Similarly, in equilibrium $C^n_{\theta}$
measures the phase fluctuations, and exhibits
a power-law, reflecting the absence of long range order in the phase. However, in the Luttinger model of interest for us here,
terms with $m > 0$ are absent from the density operator, which is given by
$\rho(x) = J_r(x) + J_l(x)$.~\cite{haldane_luttinger_liquid}}
Using exactly the same methods as above, we
find (for $L \to \infty$):
\begin{align}
\frac{C^m_{\phi} (x,t)}{C^{(0,m)}_{\phi}(x)} =
\left|\left( \frac{R_0}{2 vt } \right)^2 \frac{x^2 - (2vt)^2}{x^2} \right|^{m^2 (K^2 - 1)/2}, \\
\frac{C^n_{\phi} (x,t)}{C^{(0,n)}_{\phi}(x)} =
\left|\left( \frac{R_0}{2 vt } \right)^2 \frac{x^2 - (2vt)^2}{x^2} \right|^{n^2 (K^{-2} - 1)/8},
\end{align}
where $C^{(0,m)}(x) =  A^{\phi}_m |R_0/x|^{2m^2}$ and $C^{(0,n)}(x) = A^{\theta}_n |R_0/x|^{n^2/2}$
are the non-interacting correlation function (where $A^{\phi}_m$ and $A^{\theta}_{n}$ are
non-universal  prefactors).  We note that the usual duality relation where $\phi \to \theta$ and
$K \to K^{-1}$, which one encounters when studying equilibrium correlation functions,~\cite{giamarchi_book_1d}
still holds for these non-equilibrium correlators.  Let us next analyze their
asymptotic properties. We consider only $C^{m}_{\phi}(x,t)$, as identical conclusions also
apply to $C^{n}_{\theta}(x,t)$ by virtue of the duality relations. For $|x| \gg 2 v t$,  we have:
\begin{align}
C^{m}_{\phi}(x,t)=  C^{(0,m)}(x)\ \left(\frac{R_0}{2 v t} \right)^{m^2 (K^2-1)}.
\end{align}
Thus, up to the time-dependent pre-factor,
correlations take the form of a non-interacting system of Fermions, $C^{(0,m)}(x)$.
However, in the opposite limit  ($|x| \ll 2 v t$),  this correlator exhibits a
non-trivial power-law:
\begin{align}
C^m_{\theta}(x,t) \simeq   \left|\frac{R_0}{x} \right|^{m^2(K^2+1)}. \label{eq:long_times_2kf}
\end{align}
Notice that this expression also describes   the infinite-time behavior, which
is controlled by an exponent equal to   $m^2(K^2+1)$, being again different
from the exponent exhibited by the
same correlator in equilibrium, which equals $2 m^2 \left[\cosh 2 \beta(0) -
\sinh 2\beta(0) \right] = 2 m^2 K$.

In order to understand why the behavior found in the correlations for $t \to \infty$ in
Eqs.~(\ref{eq:long_times_correlator_zero_T},\ref{eq:long_times_2kf}),
also holds for  $|x| \ll 2 vt$, let us
consider the initial state at zero temperature,
$\rho_0 =|0\rangle \langle 0|$.~\cite{calabrese_quench_CFT,calabrese_quench_CFT_long}
As mentioned above, this is a rather complicated excited state of the Hamiltonian that
performs the time-evolution, $H_\text{f} = H_{\rm LM}$. This means that,
initially, there are a large number of excitations of
$H_\text{f}$, namely, phonons with dispersion $\omega(q) = v(q)
|q|$. The distribution of the phonons $\langle b^{\dag}(q) b(q)
\rangle_0 = \sinh^2 \beta(q)$ is time-independent and peaked at $q = 0$. Thus,
within the approximation where $v(q) \simeq v(0) = v$,  the
excitations  propagate between two given  points with  velocity
$v$. Thus, if we consider the
correlations of two points $A$ and $B$ separated a distance $|x|$,
the nature of the correlation at a give time $t$ depends on wether the excitations
found initially at, say,  point $A$,  have  been able to reach point $B$
or not. This is not the case if $|x| >  2
vt$, and thus correlations retain essentially the properties they had
in the initial state. Thus, up to a time-dependent prefactor,
$C_{\psi_r}(x,t) \propto C^{(0)}_{\psi_r}(x)$. However, if the
two points have been able to `talk to each other' through the
excitations present in the initial state, then correlations will
be qualitatively different. This happens for $t = t_0$, when the
phonons propagating from $A$ meet the phonons traveling from point $B$,
that is for $x - vt_0 = vt_0$, or $t_0 = x/2v$ (we assume $x > 0$ without
loss of generality). Thus, for
given separation $x$ and time $t$, there is a length scale $2 vt$, which
marks the transition between two different regimes in the
correlations. In the instantaneous momentum distribution, this
reflects itself in a crossover as a function of time
from a momentum distribution $n(p)$ exhibiting a discontinuous
Fermi liquid-like behavior,  which is valid
\emph{i.e.} for $|p| \ll (2vt)^{-1}$) to a power-law
behavior of the form $ \sim |p R_0|^{\gamma^2} {\rm sgn}(p)$, which
applies for $|p| \gg (2
vt)^{-1}$ but $|p| \ll R^{-1}_0$ (for $|p| \gg R^{-1}_0$ we
recover the free particle behavior corresponding to the Fermi-Dirac
distribution function at $T = 0$). In the $t \to \infty$ limit, by
using the regularization scheme described above, the asymptotic
momentum distribution at zero temperature  can be
obtained  with the help of tables.\cite{gradshteyn80_tables} The resulting formula behaves as the
non-interacting Fermi-Dirac distribution for $|p| \gg R^{-1}_0$, whereas
for $|p| \gg R^{-1}_{0}$ it describes a non-Fermi liquid-like steady state:
\begin{multline}
f(p, t \to +\infty)=\frac{1}{2}-\frac{pR_0}{2}
\Big[K_{\frac{\gamma^2-1}{2}}(|p R_0|){\cal L}_{\frac{\gamma^2-3}{2}}(|pR_0|)\\
+K_{\frac{\gamma^2-3}{2}}(|pR_0|){\cal
L}_{\frac{\gamma^2-1}{2}}(|p R_0|)\Big],
\end{multline}
where $K_{\nu}(z)$,   ${\cal L}_{\nu}(z)$ are the modified Bessel
and Struve functions,\cite{gradshteyn80_tables} respectively. This
expression yields a power law for $|p R_0|\ll 1$, where
$n(p,t\to+\infty)\simeq \frac{1}{2}-\text{const.}\times
(|pR_0|)^{\gamma^2} \, \sign(p)$. Note that the momentum
distribution $n(p =0,t) = \frac{1}{2}$, which is given by the
invariance of the LM under  particle-hole symmetry
$\psi_{\alpha}(p) \to \psi^{\dag}_{\alpha}(-p)$.

\begin{center}
\begin{figure}
\includegraphics[width=\figwidth]{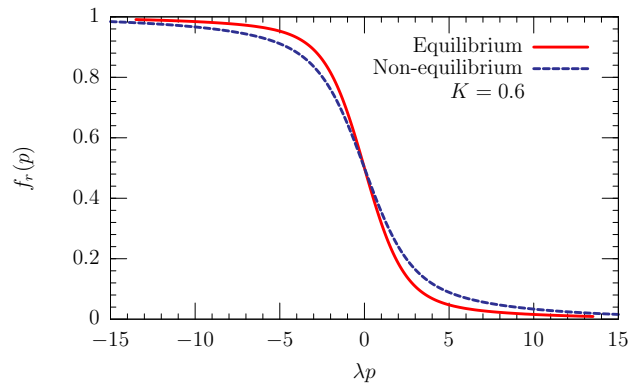}
\caption{Infinite time limit of the  momentum distribution vs.
$\lambda p$ ($\lambda = \hbar v_F /T$ is the
thermal correlation length in the initial state) for a non-interacing Luttinger model at finite
temperature $T$ that is quenched  into an interacting state with repulsive interactions
(corresonding to a Luttinger parameter $K=0.6$).}\label{fig:mom_dist_small}
\end{figure}
\end{center}

\begin{center}
\begin{figure}
\includegraphics[width=\figwidth]{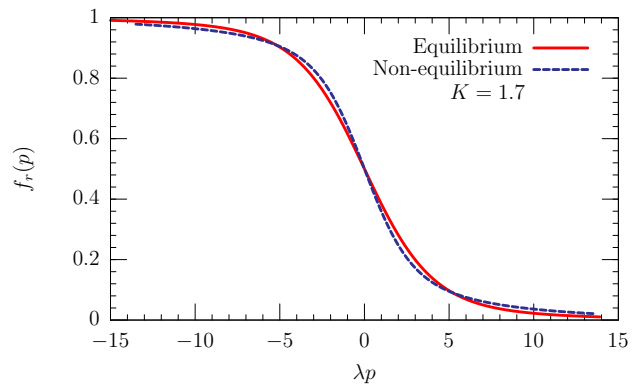}
\caption{Infinite time limit of the  momentum distribution vs.
$\lambda p$ ($\lambda = \hbar v_F /T$ is the
thermal correlation length in the initial state) for a non-interacing Luttinger model at finite
temperature $T$ that is quenched  into an interacting state with attractive interactions
(corresonding to a Luttinger parameter $K=1.7$).}\label{fig:mom_dist_large}
\end{figure}
\end{center}

Let us finally present  the generalization
of the above results for the one-particle density matrix
to finite temperatures, $T > 0$. For $T \ll \hbar v_F R_0$ (but $T \gg \Delta_0 = 2\pi \hbar v/L$,
so that we can neglect finite-size effects and effective take
the thermodynamic limit) $C_{\psi_r}(x,t)$ takes the
following asymptotic  form:
\begin{multline}
C_{\psi_r}(x,t>0|T)=C_{\psi_r}^{(0)}(x|T)\left\vert\frac{\pi  R_0/\lambda}
{dh(x|T)}\right\vert^{\gamma^2}\\
\times\left\vert\frac{dh(x-2vt|T)dh(x+2vt|T)}{dh(2vt|T)dh(-2vt|T)}
\right\vert^{\gamma^2/2},\label{eq:corr_func_fin_T}
\end{multline}
where $C^{(0)}_{\psi_r}(x|T)$ and $dh(x|T)$ can be obtained from
$C^{(0)}_{\psi_r}(x|L)$ and $d(x|L)$ by replacing $L\sin(\pi x/L)/\pi$ by
$\lambda \sinh(\pi x/\lambda)$, where $\lambda=\hbar v_F/T$ is the
thermal correlation length. At long times, $h_r(x,t|T)$ reduces to
\begin{equation}
h_r(x)=\left| \frac{\pi R_0/\lambda}{ \sinh
\left( \pi x/ \lambda \right)}
\right|^{\gamma^2}.\label{eq:factor_h_fin_T}
\end{equation}
Therefore we again find that $C_{\psi_r}(x,t\to\infty|T)$ has a form similar to the
the equilibrium correlation function at finite temperature with a different exponent
controlling the asymptotic exponential decay of correlations. Notice that the exponential
decay the correlations for $t > 0$ is a direct consequence of the fact that the
initial state has a characteristic correlation length, the thermal correlation length $\lambda$.\footnote{We shall
encounter a similar situation in Sect.~\ref{sec:sG} when analyzing quenches at $T = 0$
from a non-critical (that is, gapped) into a critical (that is, gapless) state. In that case, the role of $\lambda$ will be played by the
correlation length of the system that is determined by the (inverse of the) energy gap in the initial state.}
The exponential decay of correlations at finite $T$ implies  that the the steady state will be reached
exponentially rapidly approached in a time of the order of $\hbar/T$.
It is also worth noting, however, that  the above expression depends parametrically on the
thermal corelation length $\lambda$, and  it is the Fermi velocity,  $v_F$, which enters in the
expression for $\lambda =  \hbar v_F/T$,
instead of the (renormalized) phonon velocity which enters in the thermal length
$\lambda_{\rm eq} = \hbar v/T$, characterizing the equilibrium correlations.
Thus, since the velocity appears
only through the definition of the thermal correlation length
$\lambda$, or, in other words, in combination with the
temperature, the change from $v$ to $v_F$ can be also understood as
an change in the  temperature scale. Furthermore, in a system with Galilean
symmetry,~\cite{giamarchi_book_1d} we have that  $v K=v_F$ and thus the parameter that
controls the temperature scale now is the Luttinger parameter $K$,
so that the  asymptotic correlations at $t\to \infty$ can be regarded as the equilibrium
correlations with a different exponent and an effective temperature, $T_\text{eff}=T/K$.
Thus, for repulsive interactions (\emph{i.e.} $K<1$) we could say that, besides modifying the exponent,
quenching the system into the interacting system increases the effective temperature,
whereas for attractive interactions (\emph{i.e.} $K>1$) the effective
temperature is reduced after the quench. This effect has an impact on
the momentum distribution at finite temperatures.  To demonstrate it,
we need to compute the Fourier transform of
$h_r(x)$. This can be  done by relating it to an integral
representation of the associated Legendre function $P_\mu^\nu(z)$,
\cite{gradshteyn80_tables} and thus the Fourier transform of $h_r(x)$
can be written as:
\begin{multline}
h_r(p)=\frac{\lambda}{\sqrt{\pi}}\left(\frac{\pi
R_0}{\lambda}\right)^{(\gamma^2+1)/2}
\frac{\left|\Gamma\left(\frac{\gamma^2}{2}+\frac{i\lambda
p}{2\pi}\right)\right|^2}{\Gamma\left(\frac{\gamma^2}{2}\right)} \\
\times P_{\frac{i\lambda
p}{2\pi}-\frac{1}{2}}^{-\frac{\gamma^2}{2}+\frac{1}{2}}\left[ -\cos \left( \frac{2\pi
R_0}{\lambda}\right)\right].\label{eq:h_factor_finite_T}
\end{multline}
Hence, the momentum distribution can be obtained by numerically evaluating
the convolution with the Fermi-Dirac distribution function (cf. Eq. (\ref{eq:Fourier_transform_convolution}))
of the above expression,  Eq.~(\ref{eq:h_factor_finite_T}).
In Figs. \ref{fig:mom_dist_small} and \ref{fig:mom_dist_large}
the momentum distribution of the interacting system in the
infinite-time limit is displayed for   a non-interacting LM that is  quenched into an interacting state
with repulsive (corresponding to $K=0.6$) and attractive (corresponding to $K=1.7$) interactions,
respectively.
\subsection{Suddenly turning-off the interactions}

Next we  briefly consider the opposite situation to the one considered above,
namely  the case where in  the initial state the fermions interact and this interaction suddenly
disappears.   The fact that the initial state is a
highly complicated state of the Hamiltonian that performs the time evolution (in this case
$H_{\rm f} = H_0$,  cf. Eq.~\ref{eq:hlm1}), implies that we cannot expect that a Fermi liquid
will emerge asymptotically at long times after the quench. Indeed, at zero temperature
a thermodynamically large system
approaches a steady state exhibiting equal-time correlations that decay algebraically in space.
However, the exponents differ again from the (non-interacting) equilibrium ones. This can
be illustrated by, \emph{e.g.} computing the following correlation functions:
\begin{align}
C^m_{\phi}(x,t)  &=  \langle e^{2im \theta(x,t)} e^{-2i m \theta(0,t)} \rangle \nonumber\\
&=  I^{m}_{\phi}(x) \: \left( \frac{R_0}{2 v_F t}\right)^{m^2(K^{-1}-K)} \nonumber\\
& \quad \times \left|\frac{x^2 - (2v_Ft)^2}{x^2}\right|^{m^2(K^{-1}-K)/2},  \\
C^n_{\theta}(x,t) &= \langle e^{in \phi(x,t)} e^{-i n \phi(0,t)} \rangle \nonumber\\
&=  I^n_{\theta}(x)  \left( \frac{R_0}{2 v_F t}\right)^{n^2(K-K^{-1})/2}
\nonumber\\ & \quad \times
\left|\frac{x^2 - (2v_Ft)^2}{x^2}\right|^{n^2(K-K^{-1})/4},
\end{align}
where
\begin{align}
I^m_{\phi}(x) &= \left|\frac{R_0}{x}\right|^{2m^2 K},\\
I^n_{\theta}(x) &=  \left|\frac{R_0}{x}\right|^{n^2/2K}
\end{align}
are the correlation functions in the initial (interacting) ground state ($A_{\theta/\phi}$
are non-universal prefactors).   We note again that the duality $\theta \to \phi$ and $K \to K^{-1}$ also
holds in this case. The correlations in the stationary state that is asymptotically approached
at long times read:
\begin{align}
\lim_{t \to +\infty} C^m_{\theta}(x,t)  &=  A_{\theta} \left| \frac{R_0}{x} \right|^{m^2(K^{-1}+K)}, \\
\lim_{t \to +\infty} C^n_{\phi}(x,t)  &=  A_{\phi} \left| \frac{R_0}{x} \right|^{n^2(K^{-1}+K)/2}.
\end{align}
However, at short times, $t \ll |x|/2v_F$, correlations look like those of in the initial state,
up to a time-dependent prefactor:
\begin{align}
C^m_{\theta}(x,t \ll |x|/2v_F )  &=  \left( \frac{R_0}{2 v_F t}\right)^{m^2(K^{-1}-K)} \:  I^{m}_{\theta}(x)   \\
C^n_{\phi}(x,t \ll |x|/2v_F)  &=     \left( \frac{R_0}{2 v_F t}\right)^{n^2(K-K^{-1})/2} \:  I^n_{\phi}(x).
\end{align}
In this case the time-dependent prefactor has also a power-law form.

\section{The sine-Gordon model}\label{sec:sG}

Let us next turn our attention towards another class of models
in one dimension whose spectrum is not necessarily always critical (\emph{i.e.} gapless) like
the case of the Luttinger model (LM) analyzed previously. This is the case of the
sine-Gordon model, which is described by the following Hamiltonian:
\begin{align}
H_\text{sG}(t) &= H_0 -  \frac{ \hbar v g(t)}{\pi a^2_0} \int dx \,
\cos 2 \phi,
\label{eq:sG}\\
H_0 &= \frac{\hbar v}{2\pi} \int  dx \, : K^{-1}\left( \partial_x
\phi\right)^2 + K \left(\partial_x \theta \right)^2 :\,\,,
\end{align}
where $:\ldots:$ stands for normal order of the operators, $a_0$ is
a short-distance cut-off,  and the phase and density fields $\theta(x)$ and
$\phi(x)$, which have been introduced in the connection with the LM studied
in the previous section, are canonically conjugated in the sense that the obey:
$[\phi(x), \partial_{x'} \theta(x')] = i \pi \delta(x-x')$. This model can be regarded
as a perturbation to the LM, which still yields an integrable model. In equilibrium, the model
is known to have two phases, which, according to the renormalization
group analysis~\cite{giamarchi_book_1d,GNT98} and for infinitesimal
and positive values of the coupling  in front of the cosine term, roughly correspond
to $K < 2$ (gapped phase) and $K \geq 2$  (gapless phase).

In order to study the non-equilibrium (quench)
dynamics, we will consider two different types of quenches:
the quench from  the gapless to the gapped phase and the reversed process,
from the gapped to the gapless . In the
first case, we assume that the \emph{dimensionless} coupling
$g(t)$ is suddenly turned on, \emph{i.e.} $g(t) = g\, \theta(-t)$.
With this choice, $H_\text{i} = H_\text{sG}(t \leq 0)$ is a
Hamiltonian whose ground state exhibits a frequency  gap, $m$, to all
excitations, whereas $H_\text{f} = H_\text{sG}(t > 0)$ has gapless
excitations. Conversely, in the second case, we consider
that $g(t)$ is suddenly turned off, \emph{i.e.}
$g(t)=g\,\theta(t)$. In this case, the ground state of
$H_\text{i}$ is gapless whereas the Hamiltonian performing the
time evolution, $H_\text{f}$, has gapped
excitations. However, although both $H_\text{i}$ and $H_\text{f}$
define integrable field theories, for a
general choice of the parameters $K$ and $g$, the quench dynamics
cannot be analyzed, in general,  by the methods discussed above. Nevertheless, in two limits,
the Luther-Emery point (which corresponds to
$K = 1$, see Sect.~\ref{sect:LE})
and in the semiclassical limit (that is, for $K\ll 1$, Sect.~\ref{sect:SC}), it is possible to study  the quench
dynamics by the methods of Sect.~\ref{sec:quadratic}. However, the statistics  of the elementary
excitations happens to be different in these two cases.

\subsection{The Luther-Emery point}\label{sect:LE}

Let us start by considering the sine-Gordon model, Eq.~(\ref{eq:sG2}),
for $K = 1$, which is the so-called Luther-Emery point. It is convenient
to introduce  rescaled density and phase  fields, which will be
denoted  as $\varphi(x) = K^{-1/2} \phi(x)$ and $\tilde{\varphi}(x)
=  K^{1/2} \theta(x)$. Thus, the Hamiltonian in Eq. (\ref{eq:sG}) becomes:
\begin{multline}
H_\text{sG}(t)= \frac{\hbar v}{2\pi} \int dx :\left(\partial_x
\varphi\right)^2 +
\left(\partial_x \tilde{\varphi}\right)^2: \\
- \frac{\hbar v g(t)}{\pi a_0^2} \int dx \, \cos  \kappa \varphi,\label{eq:sG2}
\end{multline}
where $\kappa = 2 \sqrt{K}$. At the Luther-Emery point $\kappa = 2$ (\emph{i.e.} $K = 1$)
and the model can be rewritten as a one-dimensional model of  \emph{massive} Dirac  fermions with  mass
by using the bosonization formula for the Fermi field operators, Eq.~(\ref{eq:bosonization_formula}). To this end, we set
$\phi_r(x) = \varphi(x) + \tilde{\varphi}(x) $ and
$\phi_l(x) =    \varphi(x) - \tilde{\varphi}(x) $. Furthermore,
for computational convenience, we  choose the Majorana fermions in
Eq.~(\ref{eq:bosonization_formula})   to be $\eta_{r} = \sigma_x$ and $\eta_{l} = i \sigma_y$.
In addition, we note that the gradient terms in Eq.~(\ref{eq:sG2}) can be written as the kinetic
energy of free massless Dirac fermions in one dimension:~\cite{haldane_luttinger_liquid,giamarchi_book_1d,GNT98}
\begin{equation}
H_{0} =  - i \hbar v \int dx  :\psi^{\dag}_r(x) \partial_x
\psi_r(x) - \psi^{\dag}_l(x) \partial_x \psi_l(x):
\end{equation}
As far as the cosine operator is concerned,  the bosonization formula,
Eq.~(\ref{eq:bosonization_formula}), implies that:
\begin{align}
\psi^{\dag}_r(x) \psi_l(x) + \psi^{\dag}_l(x) \psi_r(x) &=
\frac{\Gamma}{\pi a_0}  \cos 2 \varphi(x) \label{eq:fermionmassterm1}\\
& = \frac{\Gamma}{\pi} :\cos 2\varphi(x) : \label{eq:fermionmassterm2}
\end{align}
where $\Gamma = i \sigma_x \sigma_y$.   This is almost the
cosine term  in the sine-Gordon model (cf. Eq.~\ref{eq:sG2})
except for the presence of the operator $\Gamma$.
However, we note that   $\Gamma^2 = 1$ and that this operator also
commutes with  $H_0$ and  with operator in the left hand-side
of  Eq.~(\ref{eq:fermionmassterm1}). The  first property implies that the eigenvalues of $\Gamma$ are $\pm 1$ whereas  the
second property implies that
$H_{\rm LE} = H_0 +  \hbar v g(t) \int dx \left[ \psi^{\dag}_r(x) \psi_l(x) + \psi^{\dag}_l(x) \psi_r(x) \right] $
and $\Gamma$ can be diagonalized simultaneously .
Upon choosing the eigen-space where $\Gamma = -1$, we obtain that
\begin{multline}
H_\text{LE}(t) =  - i \hbar v \int dx :\psi^{\dag}_r(x)
\partial_x \psi_r(x) - \psi^{\dag}_l(x) \partial_x \psi_l(x): \\
+ \hbar v g(t) \int \left[ \psi^{\dag}_r(x) \psi_l(x) +
\psi^{\dag}_l(x) \psi_r(x) \right], \label{eq:LEham}
\end{multline}
is equivalent to Eq.~(\ref{eq:sG2}) when $\kappa = 2$.

To gain some insight into the phases described by the sG model, 
let us  first consider the Luther-Emery Hamiltonian, $H_\text{LE}$
in two (time-independent) situations:  \emph{i}) $g(t) = 0$ (the gapless free
fermion phase, which  coincides with the LM for $K = 1$),  and  \emph{ii}) $g(t) = g > 0 $, \emph{i.e.}
a time-independent constant (which corresponds to the gapped phase).  In order
to diagonalize the Hamiltonian in these cases, it is convenient  to work in
Fourier space and  write the fermion field operator as:
\begin{equation}
\psi_{\alpha}(x) = \frac{1}{L} \sum_p  e^{-a_0 |p|} \, e^{i p x} \,
\psi_{\alpha}(p). \label{eq:fermifourier}
\end{equation}
where $\alpha=r,l$. The limit where the cut-off  $a_0 \to 0^{+}$ should be formally
taken at the end of the calculations, but in some cases we shall not do it
in order to regularize certain short-distance divergences of the model.
It is also useful to introduce  a spinor whose components are the right and
left moving fields, and which will make the notation more compact:
\begin{align}
\Psi(p) = \begin{bmatrix} \psi_r(p) \\ \psi_l(p)
\end{bmatrix}, \quad \mathcal{H}(p) =  \begin{bmatrix}\hbar \omega_0(p) & 0\\
0 & -\hbar \omega_0(p),
\end{bmatrix},
\end{align}
Thus the Hamiltonian for the gapless phase, $H_0$, reads:
\begin{equation}
H_0 = \sum_p:\Psi^\dagger(p)\cdot\mathcal{H}_0(p)\cdot\Psi(p):\,, \label{eq:h0le}
\end{equation}
where $\omega_0(p)=vp$ is the fermion dispersion. However, the Hamiltonian
of the gapped phase, corresponding to $g(t) = g > 0$,  $H$,   is not diagonal
in terms of the right and left moving Fermi fields. In the compact spinor notation
it reads:
\begin{align}
H =  \sum_p :\Psi^{\dag}(p) \cdot \mathcal{H}(p)
\cdot \Psi(p):,\label{eq:H_fermionic}
\end{align}
where
\begin{align}
\mathcal{H}(p) =  \begin{bmatrix}\hbar \omega_0(p) &
\hbar m\\
\hbar m & -\hbar \omega_0(p)
\end{bmatrix},\label{eq:H_fermionic_matrix}
\end{align}
being $m = v g$. Nevertheless, $H$ can be rendered diagonal
by means of the  following  unitary transformation:
\begin{equation}
\tilde{\Psi}(p) = \begin{bmatrix} \psi_{c}(p) \\
\psi_v(p)
\end{bmatrix} =  \begin{bmatrix} \cos \theta(p) &
\sin \theta(p) \\ -\sin \theta(p) & \cos \theta(p)
\end{bmatrix} \begin{bmatrix}\psi_r(p) \\ \psi_l(p)
\end{bmatrix}, \label{eq:trans}
\end{equation}
being
\begin{equation}
\tan2\theta(p) = \frac{m}{\omega_0(p)}.
\end{equation}
Thus the Hamiltonian of the gapped phase, in diagonal form, reads (we drop an
unimportant  constant that amounts to the ground state energy):
\begin{equation}
H = \sum_p \hbar \omega(p) \left[ :\psi^{\dag}_c(p) \psi_c(p) -
\psi^{\dag}_v(p) \psi_v(p): \right]. \label{hgapped}
\end{equation}
where $\omega(p) = \sqrt{\omega_0(p)^2 + m^2}$. We associate
$\psi^{\dag}_v(p)$ ($\psi^{\dag}_c(p)$) with the creation operator
for particles in the valence (conduction) band.

Before considering quantum quenches,  let us briefly discuss some of the
the properties of the ground states of the Hamiltonians
$H_0$ and $H$. In what follows, these states will be
denoted  as  $| \Phi_0\rangle$ and $| \Phi \rangle$, respectively.
As mentioned above, the spectrum of $H_0$  is gapless, and the fermion
occupancies in the ground state $|\Phi_0\rangle$ are:
\begin{align}
n_r(p) &= \langle \Phi_0| \psi^{\dag}_r(p) \psi_r(p) |
\Phi_0 \rangle = \theta(-p), \label{eq:gs11}  \\
n_l(p)  &= \langle \Phi_0| \psi^{\dag}_l(p) \psi_l(p) |
\Phi_0 \rangle = \theta(p). \label{eq:gs12}
\end{align}
That is, all single-particle levels with negative momentum are filled
out, as described in Sect.~\ref{sec:LM} (recall that $H_0$ is just the
Luttinger model with $K = 1$, \emph{i.e.} $g_2 = g_4 = 0$).
However, $H$ has a gapped spectrum and, therefore, when
constructing the ground state, $|\Phi\rangle$, only the levels in the
valence band  (which have negative energy) are filled,
whereas the levels in the conduction band remain
empty:
\begin{align}
n_v(p) &= \langle \Phi| \psi^{\dag}_v(p) \psi_v(p) |
\Phi\rangle = 1 \label{eq:occv},\\ n_c(p) &= \langle \Phi|
\psi^{\dag}_c(p) \psi_c(p) | \Phi \rangle = 0. \label{eq:occc}
\end{align}

\subsubsection{Quench from the gapped to the gapless phase}

The first situation we shall consider is when $g(t) = g\:
\theta(-t)$ in Eq.~(\ref{eq:LEham}), so that the spectrum of the Hamiltonian abruptly changes
from gapped to gapless (\emph{i.e.} quantum critical).
As we did in Sect.~\ref{sec:LM}, we denote
$H_\text{i}=H_\text{LE}(t < 0) = H$ and $H_\text{f}=H_\text{LE}(t > 0) = H_0$.
Although the expressions presented below can be computed  for finite temperature, $T > 0$,
where the initial state corresponds to $\rho_\text{i} = e^{-H_\text{i}/T}/Z_\text{i}$,
we shall restrict ourselves to the zero temperature case, where
the initial state $\rho_{\rm i} = |\Phi\rangle \langle \Phi|$.
Notice in this state, $\langle \Phi |  \cos 2 \phi(x) |\Phi \rangle =
\langle \Phi |  \cos 2 \varphi(x) |\Phi \rangle =
{\rm Re}  \, \langle \Phi|  e^{-2 i\varphi(x)}| \Phi \rangle  =  - \langle \psi^{\dag}_{r}(x) \psi_{l}(x) \rangle \neq  0$ (the minus sign
stems from the eigenvalue of the operator $\Gamma  = \eta_r \eta_l$), whereas in the
ground state of $H_0$ the expectation value of the same operator vanishes.
Therefore,  it behaves like an order parameter in equilibrium,  and we
 can expect that it exhibits interesting dynamics out of equilibrium. Indeed,
\begin{align}
\langle  e^{-2 i \varphi(x,t)} \rangle &= - \frac{1}{L} \sum_p
\langle \psi^{\dag}_r(p,t)
\psi_l(p,t)\rangle \\
&=   - \frac{1}{L}  \sum_p e^{-2i\omega_0(p)t} \, \sin 2\theta(p),
\end{align}
where, in the last expression, we have already
taken the $T \to 0$ limit  and set $\langle
\psi^{\dag}_r(p) \psi_l(p)\rangle = -\frac{1}{2} \sin 2\theta(p)$,
as it follows from Eqs.~(\ref{eq:trans},\ref{eq:occv},\ref{eq:occc}).
The above expression can be readily evaluated by recalling that
 $\sin 2\theta(p) = m/\omega(p)$, which
yields, in the $L \to \infty$ limit,
\begin{align}
\langle  e^{-2 i \varphi(x,t)} \rangle  &=  -m \int^{+\infty}_{0}
\frac{dp}{\pi} \frac{\cos2 \omega_0(p) t}{\sqrt{\omega_0^2(p) +
m^2}} \\
&= \left(\frac{m}{2\pi v} \right) K_0\left(2m t\right) \simeq
\frac{1}{4} \sqrt{\frac{m}{\pi t}} e^{-2 m t},
\end{align}
where $K_0$ is the modified Bessel function.
Thus we see that the `order parameter' $\langle \cos 2 \varphi(0,t)\rangle$ decays
exponentially at long times at $T = 0$. The decay rate is proportional to the
gap between the ground state (the initial state) and the  first excited state of the
initial Hamiltonian $H_{\rm i} = H$. The existence of this gap means, in particular,
that correlations in the initial state between  degrees of freedom of the system are exponentially suppressed
beyond a distance of the order $\xi_c \approx v/m$. Since the system is quenched into a situation where the
Hamiltonian performing the time evolution is critical, that is, characterized
by excitations that propagate along light cones $x \pm v t$, the light-cone
argument~\cite{calabrese_quench_CFT,calabrese_quench_CFT_long}
discussed in previous sections applies and translates the correlation length scale  in the initial state  into an exponential
decay in time of the order parameter.  The exponential decay found in the present case
(a quench from a gapped to a gapless or critical system) is also found  in the semiclassical approximation
to the sine-Gordon model (see Sect.~\ref{sect:SC} below),  and it is  in agreement  with the
results of Calabrese and Cardy~\cite{calabrese_quench_CFT,calabrese_quench_CFT_long}
obtained using  a mapping to boundary conformal field theory (BCFT).

Next we consider the (equal-time) two-point correlation function
of the same object:
\begin{multline}
{\cal G}(x,t) = \langle e^{-2i \varphi(x,t)} e^{2i \varphi(0,t)}\rangle =
\frac{1}{L^2} \sum_{p_1,p_2,p_3,p_4} e^{i(p_1-p_2)x}\\ \times
e^{-i[\omega_0(p_1) + \omega_0(p_2) - \omega_0(p_3) -
\omega_0(p_4)]t}\\
\times \langle \psi^{\dag}_r(p_1) \psi_l(p_2) \psi^{\dag}_l(p_3)
\psi_r(p_4)\rangle. \label{eq:twopointle2}
\end{multline}
Applying Wick's theorem, there are three different contractions of the
above four fermion expectation value, which can be evaluated using
Eqs.~(\ref{eq:trans},\ref{eq:occv},\ref{eq:occc}).This yields:
\begin{align}
\langle \psi^{\dag}_r(p) \psi^{\dag}_l(p)\rangle = \langle
\psi_r(p) \psi_l(p)\rangle &= 0\\
\langle \psi^{\dag}_l(p) \psi_r(p) \rangle = \langle \psi^{\dag}_r(p)
\psi_l(p) \rangle &= - \frac{1}{2} \sin 2\theta(p), \\
\langle \psi^{\dag}_r(p) \psi_r(p) \rangle &= \sin^2 \theta(p), \\
\langle \psi_l(p) \psi^{\dag}_l(p) \rangle &= \cos^2 \theta(p).
\end{align}
Hence, for $x \neq 0$, we obtain
\begin{multline}
{\cal G}(x,t) = \left(\frac{m}{2\pi v}\right)^2
\left(\left[K_0\left( 2 m t\right) \right]^2 +
\left[K_1\left(\frac{m |x|}{v}\right)\right]^2\right). \label{eq:twopointle3}
\end{multline}
Let us examine the behavior of this correlation function in the
asymptotic limit where $|x| \gg \xi_c \approx v/m$ and $2v t \gg \xi_c$.
Since the Bessel functions decay exponentially for large values of their arguments,
the asymptotic behavior depends
on whether $t < |x|/2v$ or $t > |x|/2v $:
\begin{align}
{\cal G}(x,t)  \simeq
\begin{cases}
 \frac{m}{16\pi v x} e^{-2m |x|/v} & t > |x|/2v,\\
 \frac{m}{32\pi v^2 t} e^{-4mt}  & t < |x|/2v.
\end{cases}
\end{align}
These results are in also agreement with those obtained by
Calabrese and Cardy for quantum quenches from a non-critical into a critical
state.\cite{calabrese_quench_CFT,calabrese_quench_CFT_long}

\subsubsection{Quench from the gapless to gapped phase}\label{sec:le_gaplesstogapped}

We next consider the reversed situation to the one discussed
in the previous subsection. In this case, we
set $g(t) = g \: \theta(t)$, \emph{i.e.} the initial state is critical and
corresponds to the ground state of $H_\text{i}=H_0$, whereas
the time evolution is performed according to $H_\text{f}=H$. To
compute the same  correlations functions as above, it is
convenient in this case to obtain the time evolution of  the
operators $\psi_r(p)$ and $\psi_l(p)$, whose action on the initial
state is known [cf. e.g. Eqs.~(\ref{eq:gs11},\ref{eq:gs12})].
Once again, we restrict ourselves to the $T = 0$
case. However, obtaining finite temperature results from the
expressions below should be rather straightforward.

We first note that the time-evolved Fermi operators can be related to the
operators at $t = 0$ by means of the following (time-dependent)
canonical transformation:
\begin{align}
\psi_r(p,t) &= e^{i H_\text{f} t} \psi_r(p) e^{-iH_\text{f} t
/\hbar} \\
&=f(p,t) \psi_r(p) + g^{*}(p,t) \psi_l(p), \label{eq:psirt}\\
\psi_l(p,t)  &= e^{i H_\text{f} t} \psi_l(p) e^{-iH_\text{f} t
/\hbar} \\
&= g^*(p,t) \psi_r(p) + f^*(p,t) \psi_l(p),\label{eq:psilt}
\end{align}
where $f(p,t) = \cos \omega(p) t - i \cos 2\theta(p) \sin
\omega(p)t$ and $g(p,t) = i \sin 2\theta(p) \sin \omega(p) t$. It
can be easily shown that the above transformation respects the
canonical anti-commutation relations characteristic of Fermi statistics.
Using Eqs.~(\ref{eq:psirt}) to (\ref{eq:psilt}),
we can now compute the decay of the order parameter. The calculation
yields:
\begin{equation}
\langle e^{-2 i \varphi(x,t)} \rangle = -\frac{2}{L} \sum_{p >0}
\Re\left[  f^*(p,t) g(p,t)  \right].
\end{equation}
In deriving the above expression we have used that $f(-p,t) = f^*(p,t)$
and $g(-p,t) = g(p,t)$, which follows from
$\cos 2\theta(-p) = - \cos 2\theta(p)$ because $\cos 2\theta(p) =
\omega_0(p)/\omega(p)$. Upon setting $ \Re \left[ f^*(p,t) g(p,t) \right] = - \cos
2 \theta(p) \sin 2\theta(p) \sin^2 \omega(p)t$ and taking
$L \to +\infty$, we find:
\begin{align}
\langle e^{-2 i \varphi(x,t)} \rangle &= -\int^{+\infty}_0
\frac{dp}{2\pi} \frac{m \omega_0(p)}{[\omega(p)]^2} e^{-a_0 p} \sin^2 \omega(p) t\\
&\simeq A(m a_0) + \frac{1}{4v t} \sin 2 mt + O(t^{-2}).\label{eq:order_param_3}
\end{align}
The first term is a non-universal constant that depends on the
energy cut-off $a_0$ introduced above (cf. Eq.~\ref{eq:fermifourier}).
Thus, we conclude that, when quenched from the critical (gapless)
phase into the gapped phase,  the order parameter exhibits
an oscillatory decay towards a  non-universal constant value, $A( m a_0)$.

Using similar methods
the  (equal-time) two-point correlation function can also be
obtained, yielding the following result:
\begin{multline}\label{eq:two_points_LE}
{\cal G}(x,t) = \langle e^{2i\varphi(x,t)}e^{-2i\varphi(0,t)}\rangle =\langle
e^{2i\varphi(x,t)}\rangle\langle e^{-2i\varphi(0,t)}\rangle\\
+\left\vert {\cal H}(x,t)+\frac{1}{L}\sum_{p>0}e^{-ipx}\right\vert ^{2}
\end{multline}
where (for $L\to \infty$) the function ${\cal H}(x,t)$ is defined as
\begin{equation}\label{eq:calh}
{\cal H}(x,t)=i\int_{0}^{\infty}\frac{dp}{2\pi}\sin px\left[1-\cos2\omega(p)t\right]\frac{m^2}{[\omega(p)]^2}\\
\end{equation}
Next we obtain the  behavior of this function for  $t\to +\infty$, in which
case, the term $\cos2\omega(p) t$  oscillates very rapidly and therefore
can be dropped.  Upon performing the momentum integral,
we obtain the following result for large $|x|$:
\begin{equation}
{\cal H}(x,t\to\infty)\approx \frac{i}{2\pi x}+\frac{4iv^2}{2\pi m^2x^3}.
\end{equation}
The first term in the above expression exactly cancels the second in the last expression in
the right hand-side of  Eq. (\ref{eq:two_points_LE}). Thus, the two-point correlation functio
 in the limit $t \to \infty$ exhibits the  following asymptotic behavior:
\begin{equation}\label{eq:two_points_LE2}
\lim_{t \to +\infty} {\cal G}(x,t) = \left[ A(m a_0) \right]^2
+\frac{4v^4}{(2\pi)^2m^4x^6}.
\end{equation}
This result is clearly different from the equilibrium
behavior of the same correlation function in the gapped phase,
where it decays exponentially to a constant.~\cite{giamarchi_book_1d,GNT98}  Indeed,
as we have just obtained,  both the order parameter and its two-point correlations
(Eqs.~\ref{eq:order_param_3} and \ref{eq:two_points_LE2}), exhibit instead an algebraic
decay to constant  values.

\subsection{The semiclassical approximation}\label{sect:SC}

A controlled approximation to the sine-Gordon model (cf. Eq.~\ref{eq:sG2})
can be obtained  in the limit where $\kappa \ll 1$  (which corresponds to
small $K$ limit in the original notation of Eq.~\ref{eq:sG}).
In this limit, we can expand the cosine term in~(\ref{eq:sG2})
about one of its minima,  \emph{e.g.}  $\varphi = 0$. Retaining
only the leading quadratic term  yields the following
\emph{quadratic} Hamiltonian for the boson field $\varphi(x)$:
\begin{multline}
H_\text{sG} \simeq H_\text{sc} = \frac{\hbar v}{2\pi} \, \int dx \left[
:\left(\partial_x \varphi(x)  \right)^2 + K \left(\partial_x
\tilde{\varphi}(x) \right)^2:\right] \\
+ \frac{\hbar v g(t) \kappa^2}{2\pi a^{2-\kappa^2/2}_0} \int dx \,
:\varphi^2(x): \label{eq:hqcf}
\end{multline}
Within this approximation, the problem of studying a quantum quench in the
sine-Gordon model becomes  akin to the general problem studied in
Sect.~\ref{sec:quadratic}.  To see this,
we first introduce the expansion in Fourier components of $\varphi(x)$,
\begin{multline}
\varphi(x) = \frac{\phi_0}{\sqrt{K}} + i \frac{\pi x}{\sqrt{K}L}
\delta N  \\
+  \frac{1}{2}\sum_{q\neq 0} \left( \frac{2\pi v}{\omega_0(q) L }\right)^{1/2}
\left[ e^{i q x} b(q) + e^{-iq x} b^{\dag}(q)
\right],\label{eq:mode_decomposition_varphi}
\end{multline}
where $\omega_0(q) = v |q|$; the $b$-operators introduced
above obey the Heisenberg algebra:
\begin{equation}
\left[ b(q), b^{\dag}(q') \right] = \delta_{q,q'},
\end{equation}
commuting otherwise. The first two terms in Eq.
(\ref{eq:mode_decomposition_varphi}) are the so-called zero-modes,
whose dynamics is only important at finite $L$. In what follows we
restrict our attention to the thermodynamic
limit ($L\to\infty$) and therefore neglect the dynamics of those
zero modes. Introducing~(\ref{eq:mode_decomposition_varphi})
into (\ref{eq:hqcf}), the Hamiltonian takes the general  form of
Eq. (\ref{eq:genham}) with $g(q,t)=m(q,t)=2 v
g(t)\kappa ^2/|q|a_0^{2-\kappa^2/2}$. Following the procedure
described in Sec.~\ref{sec:quadratic}, this Hamiltonian can be
diagonalized by means of the canonical transformation of
Eq. (\ref{eq:bogol}). Introducing $m^2=4 g
v^2\kappa^2/a_0^{2-\kappa^2/2}$, which plays the role of
the gap in the frequency spectrum of the (gapped) Hamiltonian, and
setting $m(q) = g(q)  =m^2/2\omega_0(q)$ in Eq.~(\ref{eq:bogol_beta})
yields:
\begin{equation}
\tanh 2\beta(q)=\frac{m^2/2}{\omega^2_0(q)+m^2/2}.
\label{eq:betaqq}
\end{equation}
The dispersion of the excitations in the gapped phase is:
\begin{equation}
\omega(q)=\sqrt{\omega_0(q)^2+m^2}. \label{eq:disperq}
\end{equation}
As we did when analyzing the sine-Gordon model at the Luther-Emery point,
in what follows, we shall consider the cases of a quench from the gapped to gapless  phase of the above
quadratic Hamiltonian,\footnote{This case was studied  earlier by
Calabrese and Cardy in Ref.~\onlinecite{calabrese_quench_CFT_long}, although not
as limit of the sine-Gordon model.}
Eq.~(\ref{eq:hqcf}), and the reverse situation,
from gapless to gapped.

\subsubsection{Quench from the  gapped to the gapless phase}\label{sec:sc_gappedtogapless}

Let us begin the situation where the initial state is the ground
state of $H_{\rm i} = H_{\rm sc}$ and at $ t= 0$ the Hamiltonian is changed
to $H_{\rm f} = H_{0}$. In this case,
the evolution of the expectation value of the
order parameter operator $e^{-2i\phi(x)} = e^{-i\kappa \varphi(x) }$
or the associated
correlation functions can be obtained from the knowledge of
the two-point (equal time) correlation function out of equilibrium
for the boson field $\varphi(x)$, \emph{i.e.}
${\cal C}(x,t) = \langle\varphi(x,t) \varphi(0,t)\rangle - \langle \varphi^2(0,t)\rangle$. To compute this
object when Hamiltonian changes from $H_{i} = H$ (gapped spectrum)
to $H_\text{f}$ (gapless spectrum),
we first insert the Fourier expansion of  $\varphi(x)$,
Eq.~(\ref{eq:mode_decomposition_varphi}), and use that the time
evolution (as dictated by $H_\text{f} = H_0$)
of the $b$-operators is  $b(q,t)=e^{-i\omega_0(q)t}\, b(q)$.
Finally, we use the Bogoliubov transformation,
Eq. (\ref{eq:bogol}), to relate the $b$-to the $a$-bosons, the
basis where $H_\text{i}$ is diagonal, and obtain
the following expression:
\begin{align}
\langle \varphi(x,t) \varphi(0,t) \rangle &= -\frac{1}{4}
\sum_{q\neq 0} \left( \frac{2\pi v}{\omega_0(q) L} \right) \Big[   \sinh 2\beta(q) \notag\\
& \times \cos \left( qx  - 2\omega_0(q) t \right)  \left(
2n_B(q) + 1 \right) \notag\\
&-  2 \cosh 2\beta(q) \cos q x \, n_B(q) \notag\\
&-e^{i q x} \sinh^2 \beta(q) - e^{-i qx} \cosh^2\beta(q)
\Big], \notag\\\label{eq:propphi}
\end{align}
where $n_\text{B}(q)= (e^{\hbar \omega(q)/T} - 1)^{-1}$
is the Bose factor (we have assumed that the initial state is given by the
density matrix $\rho_i = e^{-H_{\rm i}/T}/Z$).

Next let us consider the behavior of the expectation value
of the order parameter. Taking into account that $\langle
e^{-2i \phi(0,t)} \rangle = \langle e^{-i \kappa \varphi(0,t)}
\rangle = e^{-\frac{\kappa^2}{2} \langle \varphi^2(0,t)\rangle}$,
we see that $\langle \varphi^2(0,t)\rangle$ must be evaluated in closed
form using Eq.~(\ref{eq:propphi}) . Before performing any
manipulation of this expression, it is convenient to subtract the constant $\langle
\varphi^2(0,0)\rangle$, which is formally infinite (\emph{i.e.} it depends
on the short distance cut-off, $a_0$). Thus,
(at $T=0$ and $L\to\infty$) we have:
\begin{multline}
\langle \varphi^2(0,t) \rangle  - \langle \varphi^2(0,0)   \rangle = \\
\frac{1}{2}\int^{+\infty}_0 \frac{d (q v)}{\omega(q)}\,
\times \left[  \left(\frac{\omega(q)}{Œ\omega_0(q)} \right)^2 - 1 \right]
\sin^2 \omega_0(q) t. \label{eq:phi2_1}
\end{multline}
Inserting the expressions for $\omega(p)$ and $\omega_0(p)$ in the
above equation, we obtain:
\begin{equation}
\langle \varphi^2(0,t) \rangle - \langle \varphi^2(0,0)   \rangle  = -f(2mt/\hbar),
\end{equation}
where $f(z)$ is defined as:
\begin{equation}
f(z)= 1+\frac{1}{2}\,G^{21}_{13}\left(\frac{z^2}{4}\left|
\begin{matrix}
&  3/2 & \\
0 &1 &1/2
\end{matrix}\right.
\right),\label{eq:fz}
\end{equation}
being $G_{13}^{21}$ the Meijer $G$
function.\cite{gradshteyn80_tables}
Using the asymptotic expansion for this function,
$f(z)\approx 1-\frac{\pi |z|}{2}$,  and hence the long-time
behavior of the order parameter is:
\begin{equation}
\langle e^{-2i\phi(0,t)}\rangle=\langle e^{2i\phi(0,0)}\rangle
e^{\frac{\kappa^2}{8}(1-\pi m t)}. \label{eq:orderparamsc}
\end{equation}

We next examine the behavior of the two-point correlation function
of the same operator,
\begin{equation}
{\cal G}(x,t) = \langle e^{2i\phi(x,t)}e^{-2i\phi(0,t)}\rangle =e^{\frac{\kappa^2}{2}  C(x,t)},
\end{equation}
where we have defined ${\cal C}(x,t) = \langle \varphi(x,t) \varphi(0,t)\rangle
- \langle \varphi^2(0,t) \rangle$. At zero temperature, with the help of  Eq. (\ref{eq:propphi}),
we find that

\begin{multline}
{\cal C}(x,t)-{\cal C}(x,0)= -\frac{m^2}{4}\int_0^\infty\frac{d (v q)}{[\omega_0(q)]^2\omega(q)}\\
\times(1-\cos qx)(1-\cos 2\omega_0(q)t),\label{eq:cc_mslmsv}
\end{multline}
where
\begin{equation}
{\cal C}(x,0)= -\frac{1}{2}\int_0^\infty\frac{d(v q)}{\omega(q)}\left(1-\cos
qx\right)e^{-q a_0}, \label{eq:cx0}
\end{equation}
being $a_0$ is the short-distance cut-off. Evaluating
the integrals:
\begin{multline}
{\cal C}(x,t)-C(x,0)=f(mx/v)+f(2mt)\\
-\frac{f[m(x/v+2t)]+f[m(x/v-2t)]}{2},
\end{multline}
where $f(z)$ has been defined in Eq.~(\ref{eq:fz}).
Thus, asymptotically (for $\max\{x/2v,t\} \gg m^{-1}$):
\begin{equation}
{\cal G}(x,t)={  e^{\kappa^2}  \cal G}(x,0) \times\begin{cases}
e^{- \kappa^2 \pi m     |x|/2v},  \text{for }t> |x|/2v \\
e^{- \kappa^2 \pi m t},  \text{for }t<|x|/2v, \label{eq:gxtsc}
\end{cases}
\end{equation}
where ${\cal G}(x,0)$ describes the correlations in the initial (gapped ground)
state, and exhibits the following asymptotic behavior:
\begin{equation}
{\cal G}(x,0)\simeq B(a_0) \left(1-\kappa^2\frac{\pi v}{m x}e^{-m|x|/v}\right)
\end{equation}
where $B(a_0)$ is a non-universal prefactor. Thus we see that
 the asymptotic form of ${\cal G}(x,t)$ (Eq.~\ref{eq:gxtsc}),
as well that of the order parameter, Eq.~(\ref{eq:orderparamsc}), have the same form as the results
found at the Luther-Emery point, and also agree
with the  results of Calabrese and
Cardy based on BCFT.~\cite{calabrese_quench_CFT,calabrese_quench_CFT_long}

\subsubsection{Quench from the gapless to the gapped phase}\label{sec:gaplesstogapped}

In this case, the system finds itself initally in the ground state
of $H_\text{i}=H_0$, and suddenly (at $t = 0$)  the Hamiltonian is changed to $H_\text{f}=H_{\rm sc}$.
For this situation convenient to obtain the evolution of the
observables from the time-dependent canonical transformation
of Eq.~(\ref{eq:solution_time}), where  $\beta(q)$
and $\omega(q)$ are given by Eq.~(\ref{eq:betaqq}) and Eq.~(\ref{eq:disperq}),
respectively. In this case,
\begin{align}
\langle \varphi(x,t) \varphi(0,t) \rangle &= \frac{1}{4}
\sum_{q\neq 0} \left( \frac{2\pi v}{\omega(q) L} \right) \Big[   \sinh 2\beta(q) \notag\\
& \times \cos \left( qx  - 2\omega(q) t \right)  \left(
2n_B(q) + 1 \right) \notag\\
&+ 2 \cosh 2\beta(q) \cos q x \, n_B(q) \notag\\
&+e^{i q x} \sinh^2 \beta(q) + e^{-i qx} \cosh^2\beta(q)
\Big], \notag\\\label{eq:propphi2}
\end{align}
where $n_\text{B}(q)= (e^{\hbar \omega_0(q)/T} - 1)^{-1}$
is the Bose factor in the  initial state (i.e. the gapless phase).

 As in the previous case, $\langle e^{-2i \phi(x)} \rangle =
e^{-\frac{\kappa^2}{2} \langle \varphi^2(0,t) \rangle}$, and using
Eq.~(\ref{eq:propphi2}), we find that:
\begin{multline}
\langle \varphi^2(0,t)\rangle - \langle \varphi^2(0,0)\rangle = \\
\frac{1}{2}\int_0^\infty\frac{d (vq)}{\omega_0(q)}
\left[ \left( \frac{\omega_0(q)}{\omega(q)} \right)^2  -1 \right]
 \sin^2\omega(q)t
\end{multline}
Note, interestingly, that this result can be obtained from Eq.~(\ref{eq:phi2_1}) by exchanging  $\omega_0(q)$ and $\omega(q)$.
However, when evaluating the integral we find that $\langle \varphi^2(0,t) \rangle  = +\infty$, for all $t > 0$,
due to the presence of infrared divergences that  are not cured by the existence  of a gap in
the spectrum of $H_{\rm f} = H_{\rm sc}$. Thus, we   conclude that  $\langle e^{-2i \phi(x)} \rangle =
e^{-\frac{\kappa^2}{2} \langle \varphi^2(0,t) \rangle}$ vanishes at all  $t > 0$.

The above result for the evolution of the order parameter seems to indicate
that the system apparently remains critical after the quench.
This is conclusion is also supported by the behavior of
the two-point correlation function of  the operator $e^{2i\phi(x,t)}$: Let ${\cal G}(x,t) =
\langle e^{2 i \phi(x,t) } e^{-2i \phi(0,t)} \rangle = e^{\kappa^2 {\cal C}(x,t)}$,
where ${\cal C}(x,t) = \langle \varphi(x,t)\varphi(0,t)\rangle - \langle \varphi^2(0,t)  \rangle$.
Using Eq.~(\ref{eq:solution_time}) and Eq.~(\ref{eq:mode_decomposition_varphi}),
we arrive at the following result (at zero temperature,  and for $L \to +\infty$):
\begin{multline}
{\cal C}(x,t)-{\cal C}(x,0)= \frac{m^2}{4}\int_0^\infty\frac{d(v q)}{\omega_0(q)\left[\omega(q)\right]^2}(1-\cos
qx)\\
\times \left(1-\cos2\omega(q)t \right) 
\label{eq:ctc0}
\end{multline}
where
\begin{equation}
{\cal C}(x,0)=-\int_0^\infty\frac{d(v q)}{2\omega_0(q)}(1-\cos
qx)e^{-q a_0}\label{eq:cc_msl_msv_inic}
\end{equation}
 To illustrate the above point about the apparent criticality of the
non-equilibrium state,we can analyze the
behavior of the two-point correlation function, ${\cal G}(x,t)$,
in two limiting cases, for $t=0$ and $t \to +\infty$.
At $t = 0$,  the correlation function, as obtained from
Eq.~(\ref{eq:cc_msl_msv_inic}), reads:
\begin{equation}
{\cal G}(x,0)= A'(a_0) \: \left(\frac{a_0}{x}\right)^{2\kappa^2},
\end{equation}
where $A'(a_0)$ depends on the short-distance cut-off $a_0$.
Thus, the correlations are power-law because the initial state is critical.
In the limit where $t\to+\infty$, the part of the integral
in Eq.~(\ref{eq:ctc0}) containing the  term $\cos 2\omega(q)t$  oscillates
very rapidly and upon integration averages to zero.
The remaining integral can be done with the help
of tables,\cite{gradshteyn80_tables} yielding:
\begin{equation}
{\cal C}(x,t\to\infty)-{\cal C}(x,0)=-\frac{\sqrt{\pi}}{2}G^{22}_{04}\left(\frac{
m^2x^2}{4v^2}\left|
\begin{matrix}
& 1 & 1 & \\
1 & 1 & 0 & 1/2
\end{matrix}\right.
\right).
\end{equation}
where $G^{22}_{04}$ is a Meijer function Using
the asymptotic behavior of the Meijer function,\cite{gradshteyn80_tables}
we obtain:
\begin{equation}
\lim_{t \to +\infty} {\cal G}(x,t)=B'(a_0) \left(\frac{2v}{mx}\right)^{\kappa^2},
\end{equation}
$B'(a_0)$ being a non-universal constant.
Thus,  although initially the system is critical and therefore
correlations at equilibrium decay as a power law with exponent $2\kappa^2$,
when the system is quenched into a gapped
phase (where equilibrium correlations exhibit an exponential
decay characterized by a correlation length $\xi_c \approx v/m$),
the correlations remain power-law, within the semiclassical
approximation.  The exponent turns out to be smaller, equal to $\kappa^2$, which
is half the exponent in the initial (gapless) state.
In other words, within
this approximation, it seems that the system keeps
memory of its initial state, and behaves as if it was critical \emph{also}
after the quench.  This behavior seems somewhat different from the results
obtained for the same type of quench at the Luther-Emery point, where both the order
parameter and the correlations for $t +\infty$ approache a constant value,  $A(ma_0)$ (unless the
non-universal amplitude $A(m a_0) =0$, which seems to require some fine-tuning).
Whether the differences found here between the Luther-Emery point and the semi-classical
approximation are  due to a break-down of the quasi-classical
approximation (which neglects the existence of solitons and anti-solitons  in the spectrum of
the sine-Gordon model), or to a qualitative change in the dynamics as
one moves away from the Luther-Emery point, it is not clear
at the moment. To clarify this issue, further investigation of this issue will be required in
the future.

\section{Long-time dynamics and the generalized Gibbs ensemble}
\label{sec:generalized}

Recently, Rigol and coworkers~\cite{rigol_generalized_gibbs_hcbosons}
observed that, at least for observables like the momentum distribution or
the ground state density, the asymptotic (long-time)
behavior of an integrable system following a quantum quench
can described by adopting the maximum entropy (also
called `subjective') approach to Statistical Mechanics,
pioneered by Jaynes.~\cite{JaynesI,JaynesII}
Within this approach, the equilibrium state of a system is
described by  a density matrix
that extremizes the von-Neumann entropy, $S=-\Tr\rho\ln\rho$,
subject to  all possible constraints provided  by the  integrals of motion of the Hamiltonian of the
system. In the case of an integrable system, if  $\{I_{m}\}$ is a set of certain (but not all of the possible)
independent integrals of motion of the system,  this procedure leads to a `generalized' Gibbs ensemble,
described by the following density matrix:
\begin{equation}
\rho_{\rm gG} = \frac{1}{Z_{\rm gG}} e^{-\sum_m \lambda_m
I_m},\label{eq:density_operator_gG}
\end{equation}
where  $Z_{\rm gG} = \Tr e^{-\sum_m \lambda_m I_m}$. The values of the Lagrange
multipliers,  $\lambda_m,$ must be determined from the condition that
\begin{equation}
\langle I_m \rangle_\text{gG} =  {\rm Tr} \left[ \rho_{0} I_m  \right]
=  \langle I_m \rangle.
\label{eq:inic_cond_T0}
\end{equation}
where $\rho_{0}$
describes the initial state of the system, and $\langle \cdots \rangle_\text{gG}$ stands for  the average
taken over the generalized Gibbs ensemble,  Eq.
(\ref{eq:density_operator_gG}). Although  $\rho_i = |\Phi(t = 0) \rangle \langle \Phi(t= 0)|$
in the case of a pure state, as was first used in Ref.~\onlinecite{rigol_generalized_gibbs_hcbosons},
nothing prevent us from taking $\rho_{0}$ to be an arbitrary mixed state  and in particular
a thermal state characterized by an absolute temperature, $T$.  In this case,  the
 Lagrange multipliers will  depend on
$T$ or any other parameter that defines the initial state.

Rigol and coworkers tested  numerically the above
conjecture by studying the quench dynamics of a 1D lattice gas of hard-core bosons
(see Ref.~\onlinecite{rigol_generalized_gibbs_hcbosons,Rigol_hc_noneq2} for
more details). The question that naturally arises then  is whether the
family of integrable models studied in this work (see Eqs. (\ref{eq:genham}), and
their fermionic equivalences  of Eq. (\ref{eq:H_fermionic}) and
(\ref{eq:H_fermionic_matrix})) relax in agreement with the
mentioned conjecture. In other words, does the average
$\langle O\rangle(t)$  at long times relax to the value
$\langle O\rangle_\text{gG} = {\rm Tr}\:  \rho_{\rm gG} \, O$, for any
of the correlation functions considered previously?
In this section we shall illustrate some of these questions
by using the Luttinger model discussed in Sect.~\ref{sec:LM}.
We shall first show for what kind of operators the generalized Gibbs
ensemble fails to reproduce their expectation values. Moreover,
by considering the correlation function of the
current operators (\emph{i.e.} $O = J_r(x)$, cf. Eq~\ref{eq:currents}),
we will illustrate why it works. Calculations of other observables and
consideration of the  other models as  can be found in the appendices.

Let us define the generalized Gibbs ensemble for the
Luttiger model (LM). Since the final Hamiltonian (in the $N = 0$ and $J = 0$ sector)
is diagonal in the $b$-boson basis, \emph{i.e.} $H_{LM} = \sum_{q\neq 0}
\hbar v(q) |q| \: b^{\dag}(q) b(q)$, a natural choice for the set of
integrals of motion is $I_m \to I(q) = n(q) = b^{\dag}(q) b(q)$
for all $q \neq 0$ (a more complete version of the ensemble should also include $N$ and $J$,
but this will not necessary as we focus on the thermodynamic limit here).
Thus, for the quench from the non-interacting to the interacting state
(cf. Sect.~\ref{sec:turningon}), where the initial state is
$|\Phi(t =0) \rangle = |0\rangle$,  the Lagrange multipliers
are  determined by Eq.~(\ref{eq:inic_cond_T0}), which yields:
\begin{equation}\label{eq:inic_condition}
 \langle I(q) \rangle_{\text{gG}} =
 \langle n(q) \rangle_{\text{gG}} = \sinh^2\beta(q)=\frac{1}{e^{\lambda(q)}-1}.
\end{equation}
Indeed, this result can be quickly established by realizing that $\rho_{\text{gG}}$
has the same form as the density matrix of a peculiar
canonical ensemble where the temperature on each eigenmode of the final Hamiltonian
depends on the wave-vector  $q$,
that is, $T(q)=\hbar v(q)|q|/\lambda(q)$. Alternatively,
one can also regard it as an ensemble where the effective Hamiltonian that defines the
Boltzmann weight is given by $H_\text{eff}/T_\text{eff}=\sum_{q\neq 0} \lambda(q)
n(q)$. However, it is worth noting that $\rho_{\rm gG}$
is diagonal in $n(q)$, and therefore it does not capture the correlations existing
in the initial state between the $q$ and $-q$ modes. Mathematically,
\begin{multline}
\langle n(q) n(-q) \rangle = \sinh^2 \beta(q) \cosh 2 \beta(q)
\neq \langle n(q) n(-q) \rangle_{\text{gG}}  \\
=  \langle n(q) \rangle_{\text{gG}}  \langle n(-q) \rangle_{\text{gG}}
= \sinh^4 \beta(q)
\end{multline}
As matter of fact, since $n(q) n(-q)$ commutes with $H$, we
conclude from the above that $\langle n(q) n(-q) \rangle$ does not relax to the
value predicted by $\rho_{\rm gG}$. Although this defect of
$\rho_{\rm gG}$ can be fixed by enlarging the set of integrals
of motion to include the set $I'(q) = n(q) n(-q)$ as well,
we shall show below by explicit calculation that this is not needed
for the calculation of the simplest correlators and observables
in the thermodynamic limit. However, one important exception to this case
are the squared fluctuations of the energy:
\begin{multline}
\sigma^2=\langle H^2 \rangle - \langle H \rangle^2 =
\sum_{p,q}\hbar\omega(p)\hbar\omega(q)\\
\times[\langle\hat{n}(p)\hat{n}(q)\rangle-\langle\hat{n}(p)\rangle\langle\hat{n}(q)\rangle]
\end{multline}
which yields  $\sigma^2 =2\sigma^2_\text{gG}=\sum_q\sinh^2
2\beta(q)\hbar^2\omega(q)^2$.  Again, since the operator
$H^2$ is conserved, $\sigma^2$ violates the relaxation
hypothesis. However, it is  tempting to argue, because that $\sigma^2$
(as well as $\langle H \rangle$) is a non-universal
property of the LM model, this violation is less problematic
than a violation in the asymptotic behavior of the correlation functions would be,
as the latter tends to be more universal. Similar results can be
obtained for the other models considered in this work.

 Let us now consider observable correlations in the LM model.
This requires that the operator ${\cal O}$ must be hermitian, which is the case for
 the current operator $J_r(x) = \partial_x \phi_r(x)/2\pi$
 but not for the field operator $\psi_r(x)$
(in this case, one has to consider the momentum distribution, as
we have done in Sect.~\ref{sec:turningon}). In the case of the current
operator, we shall study the following two-time correlation function
(no time ordering is implied):
\begin{equation}
C_{J_r}(x,t,\tau)=\left\langle J_r(x,t+\tau/2)J_r(0,t-\tau/2)\right\rangle_T.
\end{equation}
where $\langle \ldots \rangle_T$ stands for average other the
 thermal ensemble described by $\rho_i = e^{-H_{\text{i}}/T}/Z_0$,
with $H_{\rm i} = H_0$ (cf. Eq.~\ref{eq:hlm1}). Using
(\ref{eq:boson_decomposition}),   and (\ref{eq:boson_modes}), we obtain
\begin{align}
C_{J_r}(x,t,\tau)&=\frac{1}{(2\pi)^2}\sum_{q>0}\left(
\frac{2\pi q}{L}\right)\: e^{-qa_0}\notag\\
\times&\Big\{e^{iqx} f(q,t+\tau/2)f^\ast(q,t-\tau/2)\left[ 1+n_B(q)\right]\notag\\
+&e^{iqx}g^\ast(q,t + \tau/2)g(q,t - \tau/2)n_B(q)\notag\\
+&e^{-iqx}f^\ast(q,t + \tau/2)f(q,t-\tau/2)n_B(q)\notag\\
+&e^{-iqx}g(q,t+\tau/2)g^\ast(q,t-\tau/2)\left[  1+n_B(q)
\right]\Big\},\label{eq:two_points_can}
\end{align}
being $n_B(q) = (e^{-\hbar \omega_0(q)/T} - 1)^{-1}$ ($\omega_0(q) = v_F |q|$)
the initial Bose distribution of modes.   In the following we
shall argue that, in the limit $t \to +\infty$ the above expression
reduces to the following correlator in the Generalized Gibbs
ensemble:
\begin{equation}
C_{J_r}^{gG}(x,\tau) = {\rm Tr} \left[ \rho_{\rm gG}(T) \, J_r(x,\tau)J_r(0,0) \right],
\label{eq:ggjr}
\end{equation}
where $\rho_{\rm gG}(T)$ is the extension  to
 an initial thermal state of the Generalized Gibbs ensemble
 introduced above (notice that since $[H_{\rm f}. I(q)] = 0$ and therefore $[H_{\rm f}, \rho_{\rm gG} ] = 0$,
 it is in principle possible to define time-dependent correlation functions on this ensemble, just
 as we defined the for standard equilibrium states ). For this (thermal) initial condition
(\ref{eq:inic_cond_T0}) fixes the values of $\lambda(q)$ which now
depend on $\beta(q)$ and the temperature:
\begin{equation}
\sinh^2 \beta(q) \left[1+n_B(q)\right]+\cosh^2\beta(q)
n_B(q)=\frac{1}{e^{\lambda(q)}-1}.\label{eq:cond}
\end{equation}
Introducing this result into the mode expansion for Eq.~(\ref{eq:ggjr}),
\begin{align}
C_{J_r}^{gG}(x,\tau)&=\frac{1}{(2\pi)^2}\sum_{q>0}\left(
\frac{2\pi q}{L}\right) e^{-qa_0}\notag\\
&\times\big\{ e^{iq\left( x-v \tau\right) }\cosh^2\beta(q)\left[
1+\langle n(q)\rangle\right] \notag\\
&+e^{iq\left(x+v\tau\right)  }\sinh^2\beta(q)\ \langle n(q)\rangle\notag \\
&+e^{-iq\left( x-v\tau\right) }\cosh^2\beta(q)\ \langle n(q)\rangle\big\} \notag\\
&+e^{-iq\left( x+v\tau\right)  }\sinh^2\beta(q)\left[ 1+\langle
n(q)\rangle\right],
\end{align}
we see that it coincides with the $t \to +\infty$ limit of (\ref{eq:two_points_can}),
where the rapidly oscillating terms that depend only on $t$ can be dropped

 Let us close this section with brief consideration of higher order
 correlation functions. In particular, if we consider computation objects like
 the four point current correlation function,
 $\langle J_r(x_1, t_1) J_r(x_2, t_2) J_r(x_3, t_3) J_r(x_2,t_4)\rangle_T$,
 we would encounter terms involving  $\langle n(q) n(-q)\rangle$, which
 are not described by the generalized Gibbs ensemble (in its simpler form
 where $I(q) = n(q)$). However, it is not hard to convince oneself that
 in the thermodynamic limit, $L \to \infty$ the contribution from such
 terms vanishes. This is because momentum conservation when computing
   $\langle J_r(x_1, t_1) J_r(x_2, t_2) J_r(x_3, t_3) J_r(x_2,t_4)\rangle_T$
 leaves (in most cases)  with independent two wave numbers (out of four) that must
 be summed over. When taking the thermodynamic limit, the $L^2$ resulting from this
 two sums exactly cancels the $L^{-2}$ factors  coming from the mode
 expansions of $J_r(x)$, thus yielding a finite contribution.
 However, terms involving $\langle n(q) n(-q)\rangle$
 require the four wave numbers to be equal, thus resulting in only one
 momentum summation, which therefore cannot cancel $L^{-2}$ factor.
 This justifies the use a Wick's theorem when computing higher order
 correlations in the thermodynamic limit, but in finite size systems,
this procedure as well as the simplest version of the generalized
Gibbs ensemble would miss the contribution stemming from
$\langle n(q) n(-q)\rangle$ correlations.

\section{Relevance to experiments}\label{sec:exp}

 As we described in the introduction, cold atom systems
is the ideal arena to study quench dynamics. This is because
they are, to a large extent, entirely isolated systems. Furthermore,
as far as one dimensional systems are concerned, there are
already a number of experimental
realizations, including experiments where quench dynamics has been
already probed.\cite{gunter_p_wave_interactions_1D_fermions,stoeferle_shaking%
_fast_tunnability,kinoshita_non_thermalization} Thus, in this section we
would like to  discuss the possible experimental relevance of
the results obtained in previous sections. As mentioned above,
this must be done with great care, as our results have been
obtained using effective field theory models that can be  regarded as
`caricatures' of the Hamiltonians that describe real experiments.
We must emphasize that the situation in the case of quantum quenches, in particular,
and of non-equilibrium dynamics, in general,  is very different from the analysis
of low-temperature phenomena in equilibrium. In the latter case, the
experimental relevance of effective field theories is well established using
renormalization-group arguments and it has been, over the years,
widely tested using a variety of numerical and also (when possible)
analytical methods. By contrast, here we travel through a
largely uncharted land,  and much needs to be studied
in order to achieve a similar level of rigor as in the equilibrium
case.  Thus, it is convenient to regard this models as `toy
models`, which can provide us valuable lessons and insights
into the non-equilibrium dynamics of strongly correlated systems.
This this cautionary remarks, we can proceed to discuss
some experimental systems to which the above results could
of some relevance.

 The results presented in Section \ref{sec:LM} were obtained for the
particular case of the exactly solvable LM.
The LM is the exactly solvable model describing the fixed point of a
general class of interacting one-dimensional models\cite{haldane_luttinger_liquid},
know as Tomonaga-Luttinger liquids. This class includes sytems such
as the one-dimensional Bose gas interacting via a Dirac-delta potential (which is
solvable by the Bethe-ansatz,~\cite{lieb_liniger_model}),  as well as many other systems of
interacting  Bose gases with repulsive interactions (such as dipolar) or Fermi gases with both
attractive and repulsive interactions.  With the caveats
of the previous section, it would be interesting to test the results obtained
using the LM in one of these systems. However, the dynamics may be
strongly modified by the fact that higher energy states will be also
excited following a quantum quench such as the one described here which
involves switching on (or off) the interactions. Such higher energy states
are not correctly described by the LM, and the situation is expected to
worsen in the case of short range interactions. Thus, one possible way around
is to study either experimentally or numerically quenches where the interaction
does not change between too different values and the interactions are longer range,
as the latter case provides us with a much more faithful realization of the LM.
One such system is a single-species dipolar 1D Fermi gas confined to
one dimension in a strongly anisotropic
trap (e.g. ~\cite{gunter_p_wave_interactions_1D_fermions}).
Since $p$-wave interactions are weak (away from a p-wave Feshbach resonace),
the dominant interaction is dipolar, which, when the dipoles
are all aligned by an external (electric or magnetic, depending on whether
the dipole is electric, like in hetero-nuclear molecules, or magnetic, like
in Chromium). When confined to 1D, the dipolar interaction between the atoms
 can be approximated by the potential:
\begin{equation}
V_\text{dip}(x,\theta)=\frac{1}{4\pi\epsilon_0}\;
\frac{D^2\lambda(\theta)}{(x^2+R_0^2)^{3/2}}
\end{equation}
where $D$ is the dipolar momentum of the atoms, $\theta$ is the
angle subtended by the direction of the atomic motion and the polarizing
field, and $\lambda(\theta)=(1-3\cos\theta)$. Since in this case
$g_2(q)=g_4(q)\propto\lambda(\theta)$, a sudden change in the
interactions can be produced by a sudden change in alignment of the field
with the direction of motion, that is, a change in $\theta$.  In particular, a  change in $\theta$
away from the value $\theta_m=\cos^{-1}(\frac{1}{3})$ would produce lead to a sudden switching
of the interactions amongst the Fermions.~\cite{cazalilla_quench_LL}
At zero temperature, the momentum distribution $f(p,t)$  (which can be probed by time of flight measurements)
following the quench into the interacting system would evolve as described in Sect.~\ref{sec:turningon}
(cf. Fig.~\ref{fig:fpt}), with the discontinuity at the Fermi level dying out as $t^{-\gamma^2}$.
However, currently atomic gases  are produced at temperatures $T \sim 10\%$ to $20\%$ of the 
Fermi energy, and this would complicate the observation of this effect.
If much lower temperatures could be reached in experiments, so that the application of
effective field theory is much more reliable,  we expect that in a time of the order of
$\hbar/T$ the quenched dipolar gas reaches a stationary state characterized by a
momentum distribution that differs from the thermal one. However, the calculations
of $f(p,t)$ presented in Sect.~\ref{sec:turningon} (cf. Figs.~\ref{fig:mom_dist_small}
and \ref{fig:mom_dist_large}) show that the differences between the non-equilibrium
and equilibrium results in the stationary state may be well below the current experimental
resolution. Alternatively, instead of measuring the momentum distribution, one can try
to determine the non-equilibrium exponents  measuring noise correlations in
the time-of-flight images~\cite{mathey_noise_correlations} or through
interferometry.~\cite{polkovnikov_interference_between_condensates}

 Finally, we shall remark that the sine-Gordon model is an effective field-theory
 description of the Mott insulator to superfluid transition (MI to SF) in 1D~\cite{haldane_effective_harmonic_fluid_approach,giamarchi_book_1d}
 or, when the field $\varphi$ is interpreted as the (relative) phase of two coupled 1D Bose gas, it  describes
 the Josephson coupling of two 1D Bose gases.~\cite{giamarchi_book_1d,Gritsev_spectroscopy_07} Thus, the
 results presented here  could be of some relevance when interpreting the evolution of such systems
 following a quantum quench. In particular, in the case of the MI  to SF transition, the order
 parameter $e^{-2i\phi(x)}$ describes the time evolution oscillatory part of the density in the system, whereas
 the correlations of the order parameter are related to the time evolution of the static structure factor.


\acknowledgments

We thank T. Giamarchi and A. Muramatsu for useful discussions.
AI gratefully acknowledges financial support from
the Swiss National Science Foundation under MaNEP and Division II, CONICET and UNLP
and  hospitality of DIPC, where part of
this work was done. MAC thanks M. Ueda for his kind
hospitality at the University of Tokyo during his visit
at the Ueda  ERATO Macroscopic Quantum Control
Project of JST (Japan),  during which parts of this manuscript
were completed. MAC also gratefully acknowledges financial
support of the  Spanish  MEC through grant No.
FIS2007-66711-C02-02 and CSIC through grant No. PIE 200760/007.

\appendix

\section{Details of the calculation of the one-particle
density matrix in the Luttinger model}\label{app:single_particle_correlations}

In this Appendix, we shall provide  the details of the calculation of non-equilibrium
one-particle density matrix:
\begin{equation}
C_{\psi_r}(x,t) = \langle e^{i H_\text{f} t/\hbar}
\psi^{\dag}_r(x) \psi_r(0) e^{-i H_\text{f} t/\hbar} \rangle,
\end{equation}
To this end, the formula (\ref{eq:bosonization_formula}) is used. In normal ordered form:
\begin{equation}
\psi_\alpha(x) = \frac{e^{- s_\alpha i\pi x/L}}{\sqrt{L}} \,
:e^{s_\alpha i\phi_\alpha(x)}: \quad,
\end{equation}
where the normal order is defined:
\begin{equation}
:e^{s_\alpha i\phi_\alpha(x)}:=e^{i\varphi_\alpha}e^{s_\alpha 2\pi
i x N_\alpha}e^{s_\alpha i\Phi_\alpha^\dagger(x)}e^{s_\alpha i
\Phi_\alpha(x)}.
\end{equation}
The boson field $\Phi_{\alpha}(x)$ is given by  Eq.~(\ref{eq:boson_modes}).
Hence,
\begin{multline}
: e^{-i\phi_r(x)}: \, :e^{+i\phi_r(0)}: \,=  e^{-2\pi i x N_r/L}
e^{\left[ \Phi_r(x), \Phi^{\dag}_r(0)\right]} \\ \times :e^{-i
\left[ \phi_r(x) - \phi_r(0) \right]}:\quad ,
\end{multline}
where we have used the identity $e^{A} e^{B} =  e^{[A,B]}\, e^{B}
e^{A}$, which holds provided $[A,B]$ is a c-number. Using that
($a_0 \to 0^+$ is the short-distance cut-off):
\begin{align}
\left[ \Phi_r(x), \Phi^{\dag}_r(0) \right] &= \sum_{q > 0} \left(
\frac{2\pi}{qL}\right)  e^{-q a_0} e^{i q x}\\
&=  - \ln \left[ 1 - e^{-2\pi a_0/L} e^{2i \pi x/L} \right],
\end{align}
we arrive at  the following expression for $C_{\psi_r}(x,t)$:
\begin{equation}
C_{\psi_r}(x,t) = G^{(0)}_r(x) \, \langle e^{-iF^{\dag}_r(x,t)}
e^{-i F_r(x,t)} \rangle,
\end{equation}
where
\begin{align}
G^{(0)}_r(x) &= \frac{i}{2L} \frac{1}{\sin\left[ \frac{\pi}{L}
\left( x + i a \right)\right]}  \\
F_r(x,t) &= e^{i Ht/\hbar} \left[ \Phi_r(x)   - \Phi_r(0) \right]
e^{-i H_\text{f} t /\hbar} \\
&= \sum_{q > 0} \left( \frac{2\pi}{qL}\right)^{1/2}\,  (e^{i qx} -
1)\\
&\qquad\qquad\times \left[ f(q,t) b(q) + g^\ast(q,t) b^{\dag}(-q)
\right].
\end{align}
To derive the last expression we have used Eq.~(\ref{eq:solution_time}).
Employing the identities $e^{A} e^{B} = e^{[A,B]/2} e^{A+ B}$ (provided
$[A,B]$ is a c-number) and that $\langle e^{D} \rangle =
e^{\langle D\rangle + \frac{1}{2} \langle \left( D - \langle D
\rangle \right)^2\rangle}$, we obtain:
\begin{equation}
C_{\psi_r}(x,t) =  G^{(0)}_r(x) \,  e^{- \langle F^{\dag}_r(x,t)
F_r(x,t) \rangle}.\label{eq:ffi}
\end{equation}
where we have used that $\langle  [F^{\dag}_r(x,t) ]^2 \rangle =
\langle F^2_r(x,t) \rangle =0$ because $ \langle [ b^{\dag}(q) ]^2
\rangle = \langle [ b(q) ]^2 \rangle =  \langle b^{\dag}(-q) b(q)
\rangle = 0$, and since the commutator $[ F^\dagger_r(x,t),
F_r(x,t)]$ is a c-number, it can be safely replaced by $\langle [
F^\dagger_r(x,t), F_r(x,t) ] \rangle$. Note that  for $t = 0$
$F_r(x,t =0)$ contains only $b(q)$ and thus the average $\langle
F^{\dag}_r(x,t) F_r(x,t) \rangle = 0$ at $ T= 0$. In Eq.~(\ref{eq:ffi})
the exponent can be expanded to yield:
\begin{multline}
\langle F^{\dag}_r(x,t) F_r(x,t) \rangle = \sum_{q > 0} \left(
\frac{2\pi}{qL}\right) e^{-q a_0} | e^{iqx} - 1|^2 \\
\times\left[ |f(q,t)|^2 n_0(q)  + |g(q,t)|^2  \left( n_0(q)
+1\right) \right],\label{eq:ff}
\end{multline}
being $n_0(q) = (e^{-\lambda |q|} - 1)^{-1}$ the
distribution of Tomonaga bosons in the initial state
(which has been assumed to be a mixed thermal  state),
and  $\lambda=\hbar v_F/T$ is the thermal correlation length.
We next evaluate explicitly the above result in
several limiting cases.

\subsection{Zero temperature and finite length}

Let us now consider the $T = 0$ limit, where $n_0(q) = 0$, and
thus, using (\ref{eq:fqt}) and (\ref{eq:gqt}), Eq. (\ref{eq:ff})
simplifies to:
\begin{multline}\label{eq:func_F}
\langle F^{\dag}_r(x,t) F_r(x,t) \rangle_{T=0} = \sum_{q > 0}
\left( \frac{2\pi}{qL}\right) e^{-qa_0}\sinh^2[2\beta(q)] \\
\times  \left( 1 - \cos q x \right) [1-\cos 2 v(q) |q| t].
\end{multline}
To make further progress, we shall assume that $\sinh 2 \beta(q) =
\gamma e^{-|q R_0|/2} $ where $R_0$ is the range of the
interaction. Furthermore, we shall replace $v(q)$ by $v =
v(0)$~\footnote{We assume implicitly that both $\beta(q)$ and
$v(q)$ are not singular at $q = 0$.}, what allows us to safely
take the limit $a _0\to 0+$. Next, in order to simplify the
computation, we introduce the quantity
\begin{equation}
{\cal E}_r(z)=\sum_{q > 0} \left( \frac{2\pi}{qL} \right) e^{-q R_0}
\cos qz,
\end{equation}
which can be readily computed to give
\begin{equation}
{\cal E}_r(z)=-\ln\left[\frac{\pi}{L}d(z+i R_0|L)\right]+\frac{\pi
R_0}{L}-\ln2,
\end{equation}
where $d(z|L)=L|\sin(\pi z/L)|/\pi$ is the \emph{cord} function.
Using this result into Eq. (\ref{eq:func_F}), yields the following
expression for the one-particle density matrix:
\begin{multline}\label{eq:final_l}
C_{\psi_r}(x,t>0|L)=G_r^{(0)}(x|L)\left\vert\frac{d(i R_0|L)}
{d(x+i R_0|L)}\right\vert^{\gamma^2}\\
\times\left\vert\frac{d(x+2vt+i R_0|L)d(x-2vt+i R_0|L)}{d(2vt+i
R_0|L)d(-2vt+i R_0|L)}\right\vert^{\gamma^2/2}.
\end{multline}
Taking into account that $R_0/L \ll 1$ and, in the scaling limit,
one recovers the result quoted  in the main text,
Eq.~(\ref{eq:green_function_LM}).

\subsection{Thermodynamic limit and finite temperature}

We next consider Eq.(\ref{eq:ffi}) for $L \to \infty$ and finite temperature, $T$.
Equation (\ref{eq:ff}) can be recast as:
\begin{multline}\label{eq:func_F_finite_T}
\langle F^{\dag}_r(x,t) F_r(x,t) \rangle_T = \langle F_r(x,t)
F_r(x,t) \rangle_{T = 0}\\ + {\cal H}(x) + {\cal G}(x,t)
\end{multline}
where we have introduced the following functions:

\begin{align}
{\cal H}(x) &= 2 \int^{\infty}_0 \frac{dq}{q} e^{-q a}\left(1-\cos qx \right)n_0(q), \\
{\cal G}(x,t) &= 2\gamma^2 \int^{\infty}_0 \frac{dq}{q} e^{-q R_0}
\left( 1 - \cos qx \right)\\ &\qquad\qquad\qquad [1-\cos (2 v q
t)] \, n_0(q),\label{eq:func_G}
\end{align}
which hold in the thermodynamic limit and upon replacing $v(q)$ by
$v = v(q =0)$ and $\sinh 2\beta(q) = \gamma e^{-|q R_0|/2}$ as we did
in the previous section. We next define the function
\begin{equation}
g(u;r) = 2 \int^{+\infty}_0 \frac{dq}{q} e^{-qr} \, \frac{(1 -
\cos qu)}{e^{\lambda p}-1}.
\end{equation}
which can be evaluated to yield: \cite{gradshteyn80_tables}
\begin{equation}
g(u;r) =
2\ln\left|\frac{\Gamma(1+\lambda^{-1}r)}{\Gamma[1+\lambda^{-1}(r+iu)]}\right|,
\end{equation}
where $\Gamma(z)$ is the Gamma function. In the limit where $r\ll u$,
and using that\cite{gradshteyn80_tables}
$\Gamma(z)\Gamma(1-z)=\pi/\sin(\pi z)$, the above expression
reduces to
\begin{equation}\label{eq:func_h}
g(u;r) = -\ln\left\vert\frac{dh(ir|T)}{dh(u+ir|T)}\right\vert
-\ln\left\vert\frac{u+ir}{r}\right\vert.
\end{equation}
In the previous expression we have defined:
\begin{equation}
dh(z|T)=\frac{\lambda}{\pi}\left\vert\sinh\left(\pi\lambda^{-1}z\right)\right\vert.
\end{equation}
Combining this result with Eqs.~(\ref{eq:func_F_finite_T})-(\ref{eq:func_G}) and (\ref{eq:ffi}),
it is seen that the second term in Eq.~(\ref{eq:func_h}) exactly cancels
the contributions from $G^{(0)}(x|L)$ and $\langle F_r(x,t)
F_r(x,t) \rangle_{T = 0}$ in the thermodynamic limit, and
therefore,
\begin{multline}
C_{\psi_r}(x,t|T) =
G^{(0)}_r(x|T)\left[\frac{dh(i R_0|T)}{dh(x+i R_0|T)}\right]^{\gamma^2}\\
\times\left[\frac{dh(x+2vt+i R_0|T)dh(x-2vt+i R_0|T)}{dh(2vt+i
R_0|T)dh(-2vt+i R_0|T)}\right]^{\gamma^2/2}. \label{eq:cpsiT}
\end{multline}
where
\begin{equation}
G^{(0)}_r(x|T) = \frac{i}{2\pi} \frac{\pi
\lambda^{-1}}{\sinh\left[ \pi\lambda^{-1}\left( x + i a_0
\right)\right]}.
\end{equation}
We note that the result of Eq.~(\ref{eq:cpsiT}) can be obtained from Eq.
(\ref{eq:final_l}) upon making the replacement $L\sin(\pi L^{-1}x)/\pi$ by $
\lambda\sinh(\pi \lambda^{-1} x)$. Taking into account that
$R_0/L$, and in the scaling limit, we retrieve the result quoted
in the main text, Eq.~(\ref{eq:corr_func_fin_T}).

\section{One-body density matrix of the Luttinger model
in the generalized Gibbs ensemble}\label{ap:green_function_gibbs}

Next we take up the calculation of the one-body density matrix in the
generalized Gibbs ensemble for the Luttinger model discussed in
Sect.~\ref{sec:generalized}. That is, we shall evaluate the expression
at $T = 0$.
\begin{equation}
C^{\rm gG}_{\psi_r}(x) =  \Tr\left[ \rho_{\rm gG} \:  \psi^{\dag}_r(x) \psi_r(0)\right]
\end{equation}
Using the bosonization identity, Eq.~(\ref{eq:bosonization_formula}),
we can write the expression as follows:
\begin{align}
C^{\rm gG}_{\psi_r}(x) &=  G^{(0)}_r(x) \: \langle :e^{-i\left[ \phi_r(x)
- \phi_r(0) \right]}: \rangle_\text{gG} \\
&= G^{(0)}_r(x)\,\langle e^{-i\tilde{F}^{\dag}_r(x)} e^{-i
\tilde{F}_r(x)}. \rangle_\text{gG}
\end{align}
Taking into account that
\begin{align}
\tilde{F}_r(x) &= \Phi_r(x) - \Phi_r(0) \\
&=  \sum_{q > 0} \left( \frac{2\pi}{qL}\right)^{1/2} e^{-q a/2}
(e^{iq x} - 1) \\
&\quad\quad\times \left[ \cosh \beta(q) a(q)  - \sinh \beta(q)
a^{\dag}(-q) \right].
\end{align}
The expression for $C^{\rm gG}_{\psi_r}(x)$ can be easily computed by
using the trick of regarding $\rho_{gG}$ as a canonical ensemble with
$q$-dependent temperature. Thus, following
the same steps as in the previous section we arrive at:
\begin{equation}
C^{\rm gG}_{\psi_r}(x) = G^{(0)}_r(x) \: e^{-\langle \tilde{F}^{\dag}_r(x)
\tilde{F}_r(x) \rangle_\text{gG}}.
\end{equation}
Given that
\begin{equation}
\langle \tilde{F}^{\dag}_r(x) \tilde{F}_r(x) \rangle_\text{gG}=
\sinh^2 2\beta \left[ \mathcal{D}_{r}(0) - \mathcal{D}_r(x) \right],
\end{equation}
where
\begin{align}
\mathcal{D}_r(x) &= \Re\left\{\sum_{q>0}\left(\frac{2\pi}{qL}\right)e^{-q R_0}e^{iqx}\right\}\\
&= - \ln \left| \sin \frac{\pi}{L} (x + i R_0) \right| - \ln 2 -
\frac{\pi R_0}{L}.
\end{align}
Hence, taking the thermodynamic limit
\begin{equation}
C^{\rm gG}_{\psi_r}(x)  = \frac{i}{2\pi (x + i a)} \left| \frac{R_0}{x}\right|^{\gamma^2}.
\end{equation}
Thus we see that one recovers the same results as $\lim_{t \to +\infty} C_{\psi_r}(x,t)$,
Eq.~(\ref{eq:cpsi}).

\section{The sine-Gordon model  and the generalized Gibbs model}

\subsection{Quench from the gapped to the phase at the Luther-Emery point}

In this case, the type evolution of the system is performed by $H_{0}$ (cf. Eq.~\ref{eq:h0le}),
which is diagonal in the operators  $n_{\alpha}(p) = \, : \psi^{\dag}_{\alpha}(p) \psi_{\alpha}(p):$ ($\alpha = r,l$).
Thus, the generalized Gibbs ensemble can defined by the following set of integrals of motion
$I_m \to I_{\alpha}(p) = n_{\alpha}(p)$. We see immediately that the fact this
ensemble is diagonal in $n^F_{\alpha}(p)$ means that  the order parameter,
$\langle e^{-2i\varphi(x)} \rangle_{\text{gG}} = \langle \psi_{r}(x) \psi_{l}(x) \rangle_{\text{gG}} = 0$,
which agrees with the $t \to +\infty$ limit of the order parameter, which was shown in Sect~\ref{sect:LE}
to exhibit an exponential decay to zero. However, the two-point
correlator of  $e^{2i \varphi(x)}$
has a non-vanishing limit for $t \to +\infty$. Thus,
our main concern here will be the calculation of the correlation function:
\begin{multline}
\langle e^{-2i\varphi(x)}e^{2i\varphi(0)}\rangle_{\text{gG}} = \langle \psi^{\dag}_r(x)
\psi_l(x) \psi^{\dag}_l(0) \psi_r(0) \rangle_{\text{gG}} \\
= \sum_{p_{1,}p_{2},p_{3},p_{4}}\frac{ e^{i(p_{1}-p_{2})x}}{L^{2}} \langle
\psi_{r}^{\dagger}(p_{1})\psi_{l}(p_{2})\psi_{l}^{\dagger}(p_{3})\psi_{r}(p_{4})\rangle_{\text{gG}}
\label{eq:twopointle}
\end{multline}
Since the ensemble is diagonal in the chirality index, $\alpha$,
as well as momentum, $p$, we evaluation of the above expression
can be carried out by noting that:
\begin{align}
\langle \psi_{\alpha}^{\dagger}(p)\psi_{\alpha}(p')\rangle_{\text{gG}}
&=\frac{\operatorname*{Tr}e^{-\sum_{p,\alpha'}\lambda_{\alpha'}(p)
I_{\alpha'}(p)}\psi_{\alpha}^{\dagger}(p)\psi_{\alpha}(p)}
{\operatorname*{Tr}e^{-\sum_{p,\alpha'}\lambda_{\alpha'}(p)I_{\alpha'}(p)}} \nonumber \\
& =  \frac{\delta_{p,p'}}{e^{\lambda_{\alpha}(p)}+1},
\end{align}
where the Lagrange multipliers $\lambda(q)$
can be related to the values of the same expectation values in the
initial states by imposing their conservation, that is,

\begin{align}
\langle \psi_{l}^{\dagger}(p)\psi_{l}(p)\rangle_{\text{gG}} &  =
\frac{1}{e^{\lambda_{l}(p)}+1} \\
&= \langle \psi_{l}^{\dagger}(p)\psi_{l}(p) \rangle
=\cos^{2}\theta({p})\\
\langle \psi_{r}^{\dagger}(p)\psi_{r}(p) \rangle_{\text{gG}}  &  =\frac{1}{e^{\lambda_{r}(p)}+1} \\
&= \langle \psi_{r}^{\dagger}(p)\psi_{r}(p)\rangle =\sin^{2}\theta({p}).
\end{align}
Hence,
\begin{multline}
\langle \psi_{r}^{\dagger}(p_{1})\psi_{l}(p_{2})\psi
_{l}^{\dagger}(p_{3})\psi_{r}(p_{4})\rangle_{\text{gG}}   =
\langle \psi_{r}^{\dagger}(p_{1})\psi_{r}(p_{4})\rangle _{\text{gG}}\\
\times \langle \psi_{l}(p_{2})\psi_{l}^{\dagger}(p_{3})\rangle_{\text{gG}}  \\
= \delta_{p_{1},p_{4}} \delta_{p_{2},p_{3}}\sin^{2}\theta(p_{1})
\left(1-\cos^{2} \theta(p_{2})\right)  \\
=\delta_{p_{1},p_{4}}\delta_{p_{2},p_{3}}\sin^{2}\theta(p_{1}) \sin^{2}\theta(p_{2}).
\end{multline}
Introducing the last expression into Eq.~(\ref{eq:twopointle}) yields:
\begin{equation}
\langle e^{-2i\varphi(x)}e^{2i\varphi(0)}\rangle_{\text{gG}} =
\left\vert \frac{1}{L} \sum_{p} e^{ipx}  \sin^2 \theta({p})\right\vert^2
\end{equation}
and using that $\sin^2\theta(p) = (1 - \cos^2 2 \theta(p))/2$ and
$\cos 2\theta(p) = \omega_0(p)/\sqrt{\omega^2_0(p)+m^2}$, we find
(for $x \neq 0$),
\begin{equation}
\langle e^{-2i\varphi(x)}e^{2i\varphi(0)}\rangle_{\text{gG}} =  \left(\frac{m}{2\pi v}\right)^2
\left[K_1\left(\frac{m |x|}{v}\right)\right]^2,
\end{equation}
which equals to the $t\to +\infty$ limit of Eq.~(\ref{eq:twopointle3}).
\subsection{Quench from the gapless to the gapped phase at the Luther-Emery point}
 In this case  the initial state is the
gapless ground state of $H_0$, Eq.~\ref{eq:h0le}, whereas the
Hamiltonian that performs the time evolution has a gap in the spectrum
and it is diagonal in the basis of the $\psi_v(p)$ and $\psi_c(p)$ Fermi operators
(cf. Eq.~\ref{hgapped}).  Therefore, the conserved quantities are
\begin{align}
I_{v}(p)  & = n_v(p) = \psi_{v}^{\dagger}(p)\psi_{v}(p), \\
I_{c}(p)  & = n_c(p) \psi_{c}^{\dagger}(p)\psi_{c}(p)
\end{align}
The associated Lagrange multipliers (at zero temperature), $\lambda_v(p)$ and $\lambda_c(p)$
can be obtained upon equating $\langle I_{v,c}(p) \rangle_{\rm gG} = \langle \Psi(0)| I_{v,c}(p)
| \Psi(0) \rangle$. This yields:
\begin{align}
\langle I_{v}(p)\rangle_{\rm gG} &=
\frac{1}{e^{\lambda_{v}(p)}+1} \\
&=\vartheta(-p)\sin^{2}\theta({p})+\vartheta(p)\cos^{2}\theta({p}) \label{eq:equality_gibbs_1}\\
 \langle I_{c}(p) \rangle_{\rm gG}  & =\frac
{1}{e^{\lambda_{c}(p)}+1} \\
 &= \vartheta(-p)\cos^{2}\theta({p})+\vartheta(p)\sin^{2}\theta({p}),
 \label{eq:equality_gibbs_2}
\end{align}
where $\vartheta(p)$ denotes the step function.
Using these expressions we next proceed to compute the
expectation values of the following observables:
\subsubsection{Order parameter}

We start by computing the order parameter,
\begin{align}
&\langle e^{-2i\varphi(x)}\rangle_{\rm gG}  =\langle\psi
_{R}^{\dagger}(x)\psi_{L}(x)\rangle\\
& =\frac{1}{2L}\sum_{p}\sin2\theta({p})\left[   \langle
I_{c}(p)\rangle_{\rm gG}  -\langle I_{v}
(p) \rangle_{\rm gG}\right],
\end{align}
and upon using Eqs. (\ref{eq:equality_gibbs_1}) and
(\ref{eq:equality_gibbs_2}),

\begin{align}
\langle e^{-2i\varphi(x)}\rangle_{\rm gG}
& =-\frac{1}{L}\sum_{p>0}\sin2\theta({p})\cos2\theta({p})\label{eq:le_opq1}\\
&  = -\int_{0}^{\infty}\frac{dp}{2\pi}\,
\frac{m \omega_{0}(p)}{\omega^2(p)} e^{-p a_0}  \\
& = A(m a_0)
\end{align}
where we have used that $\cos2\theta_{-p}=-\cos2\theta_{p}$;
$A(m a_0)$ is the non-universal constant introduced
in Sect.~\ref{sec:gaplesstogapped}. This result agrees with
the one obtained in Sect.~\ref{sec:gaplesstogapped} for
the order parameter in the limit $t \to +\infty$.

\subsubsection{Two-point correlation function}

 We next consider the two-point correlator of the order
parameter, namely
\begin{equation}
\begin{split}
\langle e^{2i\varphi(x)}e^{2i\varphi(0)}\rangle_{\rm gG} &=
\frac{1}{L^{2}}\sum_{p_{1,}p_{2},p_{3},p_{4}}e^{i(p_{1}-p_{2})x} \langle\psi
_{r}^{\dagger}(p_{1})\psi_{l}(p_{2}) \\
&\times \psi_{l}^{\dagger}(p_{3})\psi_{r}
(p_{4})\rangle_{\rm gG}. \label{eq:le_gG2pt}
\end{split}
\end{equation}
The calculation of the  average in this case is a bit more  involved,
but it can be performed by resorting to a factorization akin to
Wick's theorem. This is applicable only in the thermodynamic limit,
as it neglects terms the four momenta of the above expectation value
coincide. These terms yield contributions of $O(1/L)$ compared
the others.  When factorizing as dictated by Wick's theorem,  the only
non-vanishing terms are:
\begin{align}
& \langle\psi_{r}^{\dagger}(p_{1})\psi_{l}(p_{2})\psi_{l}^{\dagger}
(p_{3})\psi_{r}(p_{4})\rangle_{\rm gG} =-\delta_{p_{1}p_{4}}\delta_{p_{2}
p_{3}}  \nonumber \\
&  \times \langle \psi_{r}^{\dagger}(p_{1})\psi_{r}(p_{4})\rangle_{\rm gG}
 \langle\psi_{l}^{\dagger}(p_{3})\psi_{l}(p_{2})\rangle_{gG}\\
& +\delta_{p_{1}p_{2}}\delta_{p_{4}p_{3}}\langle\psi_{r}^{\dagger
}(p_{1})\psi_{l}(p_{2})\rangle_{\rm gG} \langle\psi_{l}^{\dagger}
(p_{3})\psi_{r}(p_{4})\rangle_{\rm gG}.
\end{align}
Upon using
\begin{align}
\langle\psi_{r}^{\dagger}(p)\psi_{r}(p)\rangle_{\rm gG}  &=
\frac{1}{2}\vartheta(p)\sin^{2}2\theta({p})  \nonumber \\
 &\quad +\vartheta(-p)\left(  1-\frac{1}{2} \sin^{2}2\theta({p})\right),\\
\langle\psi_{l}^{\dagger}(p)\psi_{l}(p)\rangle_{\rm gG} & =
\frac{1}{2}\vartheta(-p)\sin^{2}2\theta({p}) \nonumber \\
& \quad + \vartheta(p)\left(  1-\frac{1}{2} \sin^{2}2\theta({p})\right), \\
\langle\psi_{l}^{\dagger}(p)\psi_{r}(p)\rangle_{\rm gG} &=
\langle\psi_{r}^{\dagger}(p)\psi_{l}(p)\rangle_{\rm gG}   \nonumber\\
& = -\frac{1}{2}\sin2\theta({p})\cos2\theta({p})\operatorname*{sgn}(p),
\end{align}
the average over the generalized Gibbs ensemble
of the four Fermi fields on the right hand-side of
Eq.~(\ref{eq:le_gG2pt}) can be computed and yields
the following expression for the two-point correlation function
(up to terms of $O(1/L^2)$):
\begin{align}
\langle e^{2i\varphi(x)}e^{-2i\varphi(0)}\rangle_{\rm gG} & =
\Bigg\vert \frac{1}{L}\sum_{p}e^{ipx}\left[  \frac{1}{2}\vartheta(p)
 \sin^{2}2\theta({p}) \nonumber \right. \\
 & + \left. \vartheta(-p)\left(  1-\frac{1}{2}\sin^{2}2\theta_{p}\right)
\right]  \Bigg\vert ^{2} \nonumber\\
& +\left\vert -\frac{1}{L}\sum_{p>0}\sin2\theta({p})
\cos2\theta({p})\right\vert ^{2} \label{eq:le_q2}
\end{align}
The last term in the above expression in just $\langle e^{2i\varphi(x)}\rangle_{\rm gG} \:
\langle e^{-2i\varphi(0)}\rangle_{\rm gG}$ (cf. Eq.~\ref{eq:le_opq1}),
whereas the first term in the right hand-side
can be written as
\begin{multline}
\left\vert \frac{1}{L}\sum_{p}e^{ipx}\left[  \theta(-p)+\frac{1}{2}\operatorname*{sgn}(
p)\sin^{2}2\theta_{p} \right]\right\vert^2  \\
= \left\vert \frac{1}{L} \sum_{p > 0} e^{-ipx} +  \frac{i}{L}\sum_{p > 0} \sin p x  \: \frac{m^2}{\omega^2(p)} \right\vert^2\\
=  \left\vert \frac{1}{L} \sum_{p > 0} e^{-ipx} + \lim_{t \to +\infty} {\cal H}(x,t) \right\vert^2
\end{multline}
and coincides with the $t \to +\infty$ limit of the second term in the
right hand-side of  Eq.~\ref{eq:two_points_LE}
in Sect.~\ref{sec:le_gaplesstogapped} (the function ${\cal H}(x,t)$
is defined in Eq.~\ref{eq:calh}).

\subsection{Quench from the gapless to the gapped phase in the semi-classical
approximation}

In this case  Hamiltonian  performing the time evolution  is gapless,
and therefore, diagonal in the $b$-operators. Hence, the conserved quantities are%
\begin{equation}
I(q)=b^{\dagger}(q)b(q)
\end{equation}
The  Lagrange multipliers of the corresponding generalized
Gibbs density matrix are fixed from the condition:
\begin{align}
\langle I(q)\rangle_{\rm gG} &=\frac{1}{e^{\lambda(q)}-1}
= \langle \Phi(0)| b^{\dagger}(q)b(q) | \Phi(0)\rangle\\
 &=\sinh^{2}\beta(q), \label{eq:lagrange_sc}
\end{align}
where $\beta(q)$ is defined by Eq.~(\ref{eq:betaqq})
Hence, using this result we next proceed to compute the
order parameter and the two-point correlation function.
We first note that the order parameter vanishes in the generalized Gibbs
ensemble since $\langle e^{-2 i \phi(x)} \rangle_{\rm gG} =
e^{- 2 i \langle  \phi^2(0) \rangle_{\rm gG} }$ and  $\langle \phi^2(0) \rangle_{\rm gG}
=  \frac{\kappa^2}{4} \langle \varphi^2(0) \rangle$ is divergent in the
$L \to +\infty$ limit (see below).
This agrees with the  result found in Sect.~\ref{sec:sc_gappedtogapless}, where it was found
that the order parameter decays exponentially in time.  Thus, in what follows
we shall be concerned with the the two-point correlation function.

\subsubsection{Two-point correlation function}

Since $\langle e^{-2 i\phi(x)} e^{2 i \phi(0)} \rangle_{\rm gG} =  e^{-\frac{\kappa^2}{2}
{\cal C}^{\rm gG}(x)}$, where $C^{gG}(x) = \langle \varphi(x) \varphi(0) \rangle_{\rm gG}
 - \langle \varphi^2(0) \rangle_{\rm gG}$.  In order to obtain this correlator, we introduce
 the Fourier expansion of $\varphi(x)$ (neglecting the zero-mode contribution),
\begin{equation}
\varphi(x)= \frac{1}{2} \sum_{q\neq0}\left(  \frac{2\pi v}{\omega_0(q) L}\right)
^{1/2}e^{iqx}\left[  b(q)+b^{\dag}(-q)\right],
\end{equation}
into the expectation value, and using (\ref{eq:lagrange_sc})  to evaluate the
averages in the generalized Gibbs ensemble,
we find that, in the thermodynamic limit,
\begin{equation}
\langle \varphi(x) \varphi(0) \rangle_{\rm gG} = \int_{0}^{\infty}\frac{d(vq)}{4 \omega_0(q)}  \: \cos qx \: \cosh 2 \beta(q), 
\end{equation}
and therefore,
\begin{align}
{\cal C}^{gG}(x) = \langle\varphi(x)\varphi(0)\rangle_{\rm gG} - \langle \varphi^2(0) \rangle_{\rm gG}, \\ 
{\cal C}^{gG}(x) = -  \int_{0}^{\infty}\frac{d ( v q)}{\omega_{0}(q)} \: \cosh 2\beta(q) \: (1-\cos  q x )  \\
= {\cal C}(x,0) -  \frac{m^2}{4} \int^{+\infty}_0 \frac{d(vq)}{\omega(q) \left[ \omega_0(q)\right]^2} (1-\cos q x) 
\end{align}
where ${\cal C}(x,0) \equiv {\cal C}(x,t=0)$ is defined in Eq.~(\ref{eq:cx0}).
Upon comparing the last result  with Eq.~(\ref{eq:cc_mslmsv})
in the limit where $t \to +\infty$, we see they are identical.

\subsection{Quench from the gapless to a gapped phase}

In this case   the Hamiltonian that performs the time evolution is gapped, whereas
the initial state is gapless.  Thus, differently from the previous case, the Hamiltonian
that performs the evolution is diagonal in the $a$-operators, and therefore, the
conserved  quantities are $I(q)=a^{\dagger}(q)a(q)$. The corresponding Lagrange
 (at zero temperature) are fixed from the condition:
\begin{align}
 \langle I(q)\rangle_{\rm gG}  & =\frac
{1}{e^{\lambda(q)}-1}=\left\langle \Phi(0) | a^{\dag}(q) a(q)| \Phi(0) \right\rangle\\
& =\sinh^{2}\beta(q),
\end{align}
where $\beta(q)$ is given by Eq.~(\ref{eq:betaqq}).

 In order to obtain the one and  two-point correlation functions
of  $e^{2 i \phi(x)} = e^{2 i \kappa \varphi(x)}$
we first need to write the $\varphi(x)$ field in terms of the $a$-operators. Upon
using the canonical transformation Eq.~(\ref{eq:bogol}):
\begin{equation}
\varphi(x) = \frac{1}{2}\sum_{q\neq 0} \left(  \frac{2\pi}{\omega(q) L} \right)^{1/2} e^{i qx} \left[a(q) + a^{\dag}(-q) \right].
\end{equation}
Hence, since $ \langle e^{-2i \phi(x)} \rangle_{\rm gG} =
\langle e^{- i \kappa \varphi(x) } \rangle_{\rm gG} = e^{ - \frac{\kappa^2}{2} \langle \varphi^2(0) \rangle_{\rm gG}}$,
and $\langle \varphi^2(0) \rangle_{\rm gG}$ is logarithmically divergent in the thermodynamic limit (see expressions below),
the find that  $\langle e^{- i \kappa \varphi(x) } \rangle_{\rm gG} = 0$. This result is in agreement with the one found in
Sect.~\ref{sec:gaplesstogapped} for the order parameter.

\subsubsection{Two-point correlation function}

 Next we consider the two-point correlation function of the same
operator, namely $\langle e^{-2 i \phi(x)} e^{2i \phi(0)} \rangle_{\rm gG} = e^{-\frac{\kappa^2}{2} {\cal C}^{\rm gG}(x)}$,
where ${\cal C}^{\rm gG}(x) = \langle \varphi(x) \varphi(0) \rangle_{\rm gG} - \langle \varphi^2(0)\rangle_{\rm gG}$.
We first obtain:
\begin{align}
\langle \varphi(x)\varphi(0)\rangle_{\rm gG}
&=  \int_{0}^{\infty}\frac{d(vq)}{4 \omega(q)}  \: \cos qx \: \cosh 2 \beta(q).
\end{align}
Hence,
\begin{align}
{\cal C}^{\rm gG}(x)  &=   - \int_{0}^{\infty}\frac{ d (vq)}{2\omega(q)}\:  \cosh \beta(q)   \left( 1- \cos qx \right)\: \\
  &= {\cal C}(x,0) + \frac{m^2}{4}\int^{+\infty}_{0} \frac{d(v q)}{\omega_0(q) [\omega(q)]^2} (1 - \cos q x) 
 \end{align}
where ${\cal C}(x,0)$ is defined in Eq.~(\ref{eq:cc_msl_msv_inic}).
The latter result agrees with Eq.~(\ref{eq:ctc0}) in the $t \to +\infty$ limit.

\bibliography{phys}

\end{document}